\documentclass[conference]{IEEEtran}
\usepackage{cite}

\ifCLASSINFOpdf
\else
\fi
\usepackage{float}
\usepackage{graphics}
\usepackage{graphicx}
\usepackage{caption}
\usepackage{epstopdf}

\usepackage{amsfonts}

\usepackage[caption=false,font=footnotesize]{subfig}
\usepackage{algorithm}  
\usepackage{algpseudocode}  
\usepackage{amsmath}  
\usepackage{multicol}
\usepackage{multirow}

\usepackage{balance}

\usepackage{color}

\def\blue#1{\textcolor{black}{#1}}
\def\org#1{\textcolor{black}{#1}}
\def\mage#1{\textcolor{black}{#1}}

\def\colorA#1{\textcolor[rgb]{0,0,0}{#1}}
\def\colorB#1{\textcolor[rgb]{0,0,0}{{#1}}}

\def\colorC#1{\textcolor[rgb]{0,0,0}{#1}}
\def\colorD#1{\textcolor[rgb]{0,0,0}{#1}}
\def\colorE#1{\textcolor[rgb]{0,0,0}{#1}}

\begin{document}
%
\title{Data Poisoning Attacks to Deep Learning Based Recommender Systems}


\author{
\IEEEauthorblockN{Hai Huang$^{1}$, Jiaming Mu$^{1}$, Neil Zhenqiang Gong$^{2}$, Qi Li$^{1}$, Bin Liu$^{3}$, Mingwei Xu$^{1}$}
\IEEEauthorblockA{
$^{1}$Institute for Network Sciences and Cyberspace \& Department of Computer Science and Technology, Tsinghua University\\
$^{1}$Beijing National Research Center for Information Science and Technology (BNRist)\\
$^{2}$Duke University \\
$^{3}$Department of Management Information Systems, West Virginia University
}
}


%


\IEEEoverridecommandlockouts
\makeatletter\def\@IEEEpubidpullup{6.5\baselineskip}\makeatother
\IEEEpubid{\parbox{\columnwidth}{
    Network and Distributed Systems Security (NDSS) Symposium 2021\\
    21-24 February 2021\\
    ISBN 1-891562-66-5\\
    https://dx.doi.org/10.14722/ndss.2021.24525\\
    www.ndss-symposium.org
}
\hspace{\columnsep}\makebox[\columnwidth]{}}

\maketitle

\begin{abstract}

Recommender systems play a crucial role in helping users to find their interested information in various web services such as Amazon, YouTube, and Google News. Various recommender systems, ranging from neighborhood-based, association-rule-based, matrix-factorization-based,
\org{to}
deep learning based, have been developed and deployed in industry. Among them, deep learning based recommender systems become increasingly popular due to their superior performance. 

In this work, we conduct the first systematic study on data poisoning attacks to deep learning based recommender systems. An attacker's goal is to manipulate a recommender system such that the attacker-chosen target items are recommended to many users.
To achieve this goal, our attack injects fake users with carefully crafted ratings to a recommender system.  
Specifically, we formulate our attack as an optimization problem, 
such that the injected ratings would maximize the number of normal users to whom the target items are recommended.
However, it is challenging to solve the optimization problem because it is a non-convex integer programming problem. To address the challenge, we develop multiple techniques to approximately solve the optimization problem. Our experimental results on \colorA{three} real-world datasets, \colorB{including small and large datasets,} show that our attack is effective and outperforms existing attacks. Moreover, we attempt to detect fake users via statistical analysis of the rating patterns of normal and fake users. Our results show that our attack is still effective and outperforms existing attacks even if such a detector is deployed. 
\end{abstract}



%

\section{Introduction}
In the era of data explosion, people often encounter information overload problems in their daily lives. For example, when they are shopping online, reading news, listening to music or watching videos, they often face challenges of  choosing their interested items from a huge number of candidates. Recommender systems help people find their interested items easily by mining historical user-item interaction data. Therefore, recommender systems have been widely used in the real world, which brings huge economic benefits.

{\colorA{Unlike the non-personalized recommender system that  recommends the same items to all users, personalized recommender system that we focus in this work uses users' historical behavior (e.g., ratings or clicks) to model their preferences and make personalized recommendations for each user~\cite{pi2018survey}.} In a typical \colorA{personalized }recommender system setting, we are given a set of users, a set of items, and  a log of the users' historical interactions (e.g., ratings) with the items, and the goal is to recommend each user a list of top ranked items based on user preferences learned from the historical interactions.}
Traditional recommender systems include \textit{neighborhood-based}~\cite{sarwar2001item}, \textit{association-rule-based}~\cite{davidson2010youtube}, \textit{matrix-factorization-based} (a.k.a \textit{latent factor model})~\cite{koren2009matrix}, and \textit{graph-based}~\cite{fouss2007random}. 
Recently, with the rapid development of  deep learning techniques, deep neural networks  have been applied to empower recommender systems~\cite{cheng2016wide,covington2016deep,he2017neural,okura2017embedding}. 
\colorA{Moreover, due to various advantages, such as nonlinear transformation and representation learning that cannot be realized by traditional techniques, deep learning is gradually becoming a technology trend in the field of recommender systems~\cite{zhang2019deep}.}

Meanwhile, various studies have shown that recommender systems are vulnerable to \emph{data poisoning attacks}~\cite{mobasher2007toward,lam2004shilling,yang2017fake,li2016data,xing2013take,fang2018poisoning, fang2020influence} \colorA{(a.k.a \emph{shilling attacks}~\cite{gunes2014shilling})}. 
Particularly, in a data poisoning attack, an attacker injects fake users with carefully crafted ratings to a recommender system such that the recommender system makes recommendations as the attacker desires, e.g., an attacker-chosen target item is recommended to many normal users. Data poisoning attacks pose severe threats to the trustworthiness of recommender systems and could manipulate Internet opinions. For instance, if an attacker manipulates a news recommender system such that a particular type of news are always recommended to users, then the attacker may be able to manipulate the users' opinions.  
However, existing data poisoning attacks are either agnostic to recommender system algorithms~\cite{mobasher2007toward,lam2004shilling} or optimized to traditional recommender system algorithms such as association-rule-based~\cite{yang2017fake}, graph-based~\cite{fang2018poisoning}, and matrix-factorization-based~\cite{li2016data,fang2020influence}. Although deep learning based recommender systems gain increasing attention and are deployed in industry, their security against data poisoning attacks is largely unknown. 

In this work, we aim to bridge this gap. Specifically, we propose data poisoning attacks that are optimized for deep learning based recommender systems. We consider an attacker's goal is to \textit{promote} a target item in a deep learning based recommender system, i.e., {an attacker-chosen target item is recommended to a large number of users}. 
To achieve this goal, an attacker injects fake users with carefully crafted ratings to the recommender system. 
As resources are limited in an attack, we assume that
the attacker can only inject a limited number of fake users 
and each fake user rates a limited number of items {(including the target item and other non-target items)} to evade trivial detection. 
The key challenge of constructing the attack is to choose the {rated} items for each fake user. To address the challenge, 
we formulate the attack as an optimization problem with an objective function of maximizing the \emph{hit ratio} of the target item, where the hit ratio of an item is the fraction of normal users to whom the item is recommended. 

However, the optimization problem is difficult to solve because of the following reasons: i) the {inputs} of the problem, i.e., data of users and items in deep learning based recommender systems, are discrete variables, and ii) the training process for a deep neural network is time-consuming, which makes it impossible for any method to require a large number of training iterations for solving the problem. Thus, we develop heuristics to approximately solve the optimization problem. Instead of directly generating the desired {rated items} for fake users, we train a surrogate model called \textit{poison model} and carefully modify it to simulate the target deep learning based recommender system. Then, we utilize this poison model to predict the rating score vector of each fake user{, and then we process the vector to assist in selecting the rated items for each fake user, so as to achieve our goal effectively.} 

{We evaluate our attack and compare it with existing  data poisoning attacks using three real-world datasets with different sizes, i.e., MovieLens-100K~\cite{harper2015movielens}, Last.fm~\cite{Cantador:RecSys2011}, and MovieLens-1M~\cite{harper2015movielens}. Our results show that our attack can effectively promote target items and significantly surpasses the baseline attacks under the white-box setting. For example, via inserting only $5\%$ of fake users, our attack can make unpopular target items recommended to about 52.6 times more normal users in the Last.fm dataset. Moreover, on the larger MovieLens-1M~\cite{harper2015movielens} dataset, our attack achieves a hit ratio of 0.0099 for random target items when injecting only $5\%$ of fake users, which is about 1.2 times of the best hit ratio achieved by the baseline attacks. We further explore the impact of partial knowledge on our poisoning attack under two different partial knowledge settings. We observe that our attack remains effective and significantly outperforms the baseline attacks in these settings. For example, when the attacker only knows $30\%$ of ratings in the original user-item rating matrix, our attack obtains a hit ratio of 0.0092 for random target items when injecting $5\%$ of fake users on the MovieLens-1M dataset, which is at least 1.3 times of the hit ratio of the baseline attacks. In addition, our attack is transferable to structure-unknown deep learning based recommender systems. In particular, even if we do not know the exact neural network architecture used by the target recommender system, our attack still makes random target items recommended to about 5.5 times more normal users when injecting $5\%$ of fake users in the MovieLens-100K dataset.  
Our results demonstrate that our attack poses a severe security threat to deep learning based recommender systems.}

Moreover, we explore detecting fake users via statistical analysis of their rating patterns and measure the attack effectiveness under such detection. The intuition behind the detection is that fake users may have rating patterns that are statistically different from those of normal users as they are generated according to specific rules. Particularly, for each user, we extract multiple features from its ratings. 
Then, we train a binary classifier to distinguish between fake and normal users based on the feature values and utilize the SVM-TIA~\cite{zhou2016svm} method to detect potential fake users. The service provider removes the detected fake users  before training the recommender system. 
Our experimental results show that such a method can effectively detect the fake users generated by existing attacks. However, the method falsely identifies a large fraction \colorB{(e.g., $30\%$)} of the fake users constructed by our attack as normal users. 
As a result, our attack is still effective and significantly outperforms  existing attacks even if such a detection method is deployed. 

The contributions of our paper are summarized as follows:
\begin{itemize}
\item We perform the first systematic study on data poisoning attacks to deep learning based recommender systems.
\item We formulate our attack as an optimization problem and develop multiple techniques to approximately solve it. 
\item We  evaluate our attack and compare it with existing ones on \colorA{three} real-world datasets.

\item We study detecting fake users via statistical analysis of their ratings and its impact on the effectiveness of data poisoning attacks.
\end{itemize}

\section{Background and Related Work}

In this section, we briefly introduce recommender systems and existing data poisoning attacks to them.

\subsection{Recommender Systems}
We consider a typical collaborative filtering based recommender system setting where we have $M$ users and $N$ items,  and we are given a record of the users' past user-item interactions $\{\langle u, i, y_{ui} \rangle\}$, where $y_{ui}$ denotes the preference of user $u$ to item $i$. The observed  user-item interactions $\{\langle u, i, y_{ui} \rangle\}$ can be represented as a user-item interaction matrix $\mathbf{Y}\in \mathbb{R}^{M\times N}$. Typically, $\mathbf{Y}$ is extremely sparse, i.e., on average each user would have  interactions with only a small portion of all the $N$ items.  We use a row vector of $\mathbf{Y}$, indicated as $\mathbf{y}_{(u)}$ (i.e., $\mathbf{y}_{(u)}=\{y_{u1},y_{u2},\ldots,y_{uN}$\}), to represent each user $u$, and a column vector of $\mathbf{Y}$, indicated as $\mathbf{y}^{(i)}$ (i.e., $\mathbf{y}^{(i)}=\{y_{1i},y_{2i},\ldots,y_{Mi}$\}) , to represent each item $i$.
Then, the task of a recommender system can be transformed into inferring a complete predicted interaction-matrix $\widehat{\mathbf{Y}}$ based on $\mathbf{Y}$, where $\widehat{y}_{ui}$ in $\widehat{\mathbf{Y}}$ denotes the predicted score of $y_{ui}$. The inferred interaction-matrix $\widehat{\mathbf{Y}}$ is then used to recommend to users a list of items that the users have not experienced yet. Specifically, if we want to recommend  $K$ items for user $u$, we select the top $K$ items that (1) they have not been rated by the user, and that (2) they have the highest predicted sores in the row vector $\widehat{\mathbf{y}}_{(u)}$ (i.e., $\widehat{\mathbf{y}}_{(u)}=\{\widehat{y}_{u1},\widehat{y}_{u2},\ldots,\widehat{y}_{uN}$\}) of $\widehat{\mathbf{Y}}$.

Depending on how to analyze the user-item interaction matrix, traditional collaborative filtering based recommender systems can be roughly divided into four categories, i.e., \textit{neighborhood-based}~\cite{sarwar2001item}, \textit{association-rule-based}~\cite{davidson2010youtube}, \textit{matrix-factorization-based} (a.k.a \textit{latent factor model})~\cite{koren2009matrix}, and \textit{graph-based}~\cite{fouss2007random}. Due to good performance and flexibility in compositing more sophisticated models, matrix factorization (MF) has become the most widely used approach among them.

More recently, with the rapid development of the deep learning techniques, deep neural networks have been applied to recommender systems and have been found to outperform traditional methods in various aspects. Deep learning based recommender systems use different neural networks structures to model user-item interactions to boost recommendation performance~\cite{zhang2019deep}.
For example,  Multilayer Perceptron (MLP)~\cite{he2017neural,hornik1989multilayer}, Autoencoder (AE)~\cite{chen2012marginalized}, Adversarial Networks (AN)~\cite{goodfellow2014generative}, and Deep Reinforcement Learning (DRL)~\cite{zhang2019deep, mnih2015human} have been applied to recommender systems to improve the recommendation accuracy.

In this paper, without loss of generality, we focus on a general deep learning based \colorE{recommender} system framework, Neural Collaborative Filtering (NCF)~\cite{he2017neural}.
NCF explores deep neural networks to model sophisticated nonlinear user-item interactions. Note that MF-based recommendation methods assume a latent factor vector to represent each user and each item, and apply a simple linear model on the user and item vectors to capture the user-item interactions. In contrast, NCF uses deep neural networks to capture nonlinear user-item interactions by passing the user and item latent factor vectors through multilayer perceptron (MLP). The output layer of NCF is the prediction of the user-item interaction $y_{ui}$. 

\begin{figure}[!t]
\centering
\includegraphics[width=0.43\textwidth]{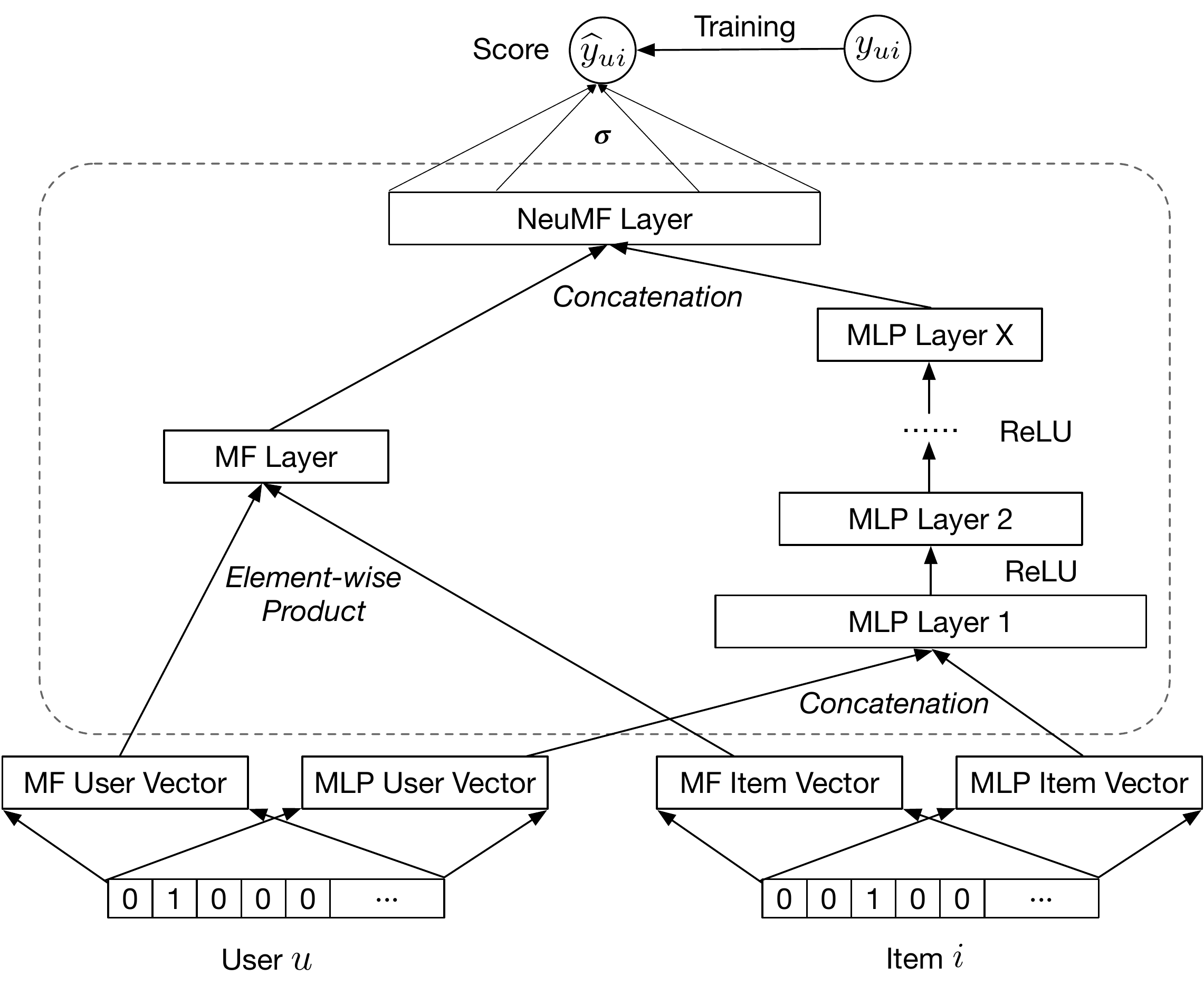}
\caption{Neural matrix factorization model (NeuMF), an instantiation of Neural Collaborative Filtering (NCF) \cite{he2017neural}.} 
\vspace{-4mm}
\label{fig:NeuMF}
\end{figure}

In particular, we consider neural matrix factorization (NeuMF)~\cite{he2017neural},  {an instantiation} of NCF, to model user-item interactions. As shown in Figure~\ref{fig:NeuMF}, \blue{NeuMF is a fusion of MF and MLP, which allows them to learn separate embeddings and then combines the two models by concatenating their last hidden layers. } \blue{The input layer consists of two binarized sparse vectors with one-hot encoding for the  user $u$ and item $i$, respectively. These sparse vectors are then separately projected into four dense latent vectors, i.e., MF user vector, MF item vector, MLP user vector, and MLP item vector, two of which are the embeddings for the user and the item in the MF model, and the others are those in the MLP model.} 
There are then two parts separately \blue{processing} latent vectors. 
One is a linear MF part, which uses a MF layer to compute the inner product of MF user vector and MF item vector, and the other is a nonlinear MLP part, which adds a standard MLP with $X$ layers on the concatenated latent vector to learn the nonlinear interaction between user $u$ and item $i$, where $X$ is the number of MLP layers and the activation function in the MLP layers is ReLU~\cite{glorot2011deep}. Finally, the \blue{last hidden layers} of MF part and MLP part are concatenated and \blue{fully connected to the output layer to predict $\widehat{y}_{ui}$}.
\blue{After training using the observed user-item interactions, this model can predict the missing entries in the original sparse interaction matrix $\mathbf{Y}$ to constitute a predicted interaction matrix $\widehat{\mathbf{Y}}$ which can be further used for constructing recommendation list for each user.}

\subsection{Attacks to Recommender Systems}

Existing studies showed that recommender systems are vulnerable to various security attacks~\cite{yang2017fake,li2016data,lam2004shilling,calandrino2011you}, which deceive a recommender system, e.g., to promote a target item and recommend it to as many users as possible. Roughly speaking, there are two categories of {such} attacks, i.e., data poisoning attacks \colorA{(a.k.a shilling attacks)}~\cite{yang2017fake,fang2018poisoning,li2016data,lam2004shilling, gunes2014shilling, dong2020trust,fang2020influence} and profile pollution attacks~\cite{xing2013take}, which compromise a recommender system at training and testing, respectively. Specifically, \blue{data poisoning attacks aim to spoof a recommender system to make attacker-desired recommendations by injecting fake users to the recommender system}, while profile pollution attacks \blue{intend} to pollute the historical behavior of normal users to manipulate the recommendations \blue{for them}.

\noindent \textbf{Data Poisoning Attacks.} 
Data poisoning attacks inject fake users to a recommender system and thereby modify the recommendation lists. Specifically, to construct a poisoning attack, the attacker first needs to register a number of fake users in a web service associated with the recommender system. Each fake user generates well-crafted rating scores for a chosen subset of items. These fake data will be included in the training dataset of the target recommender system and then poisons the training \blue{process}. 
\blue{According to whether data poisoning attacks are focused on a specific type of recommender system, we can divide them into two categories: \textit{algorithm-agnostic} and \textit{algorithm-specific}. The former (e.g., \colorA{types of shilling attacks like} {\em random attacks}~\cite{kapoor2017review, mobasher2007attacks} and {\em bandwagon attacks}~\cite{kapoor2017review, o2005recommender}) does not consider the algorithm used by the recommender system and therefore often has limited effectiveness. For instance, random attacks just choose rated items at random from the whole item set for fake users, and bandwagon attacks tend to select certain items with high popularity in the dataset for fake users. The algorithm-specific data poisoning attacks are optimized to a specific type of recommender systems and have been developed for graph-based recommender systems~\cite{fang2018poisoning}, association-rule-based recommender \blue{systems}~\cite{yang2017fake}, matrix-factorization-based recommender systems~\cite{li2016data, fang2020influence}, {and neighborhood-based recommender systems~\cite{chen2020data}}. As these attacks are optimized, they often are more effective. 
However, there is no study on algorithm-specific data poisoning attacks to deep learning based recommender systems. We bridge this gap in this paper.}

\noindent \textbf{Profile Pollution Attacks.} The key idea of profile pollution attacks is to pollute a user’s profile (e.g., historical behavior) via cross-site request forgery (CSRF)~\cite{zeller2008cross}. For instance, Xing et.al.~\cite{xing2013take} proposed profile pollution attacks to recommender systems in web services, e.g., YouTube, Amazon, and Google. Their study shows that all these services are vulnerable to their attacks. However, profile pollution attacks have two key limitations: i) profile pollution attacks rely on CSRF, which makes it hard to perform the attacks at a large scale, and ii) profile pollution attacks can not be applied to item-to-item recommender systems because the attackers are not able to pollute the profile of an item~\cite{yang2017fake}.

\section{Problem Formulation}\label{sec:problem}

In this section, we first present our threat model and then we  formulate our poisoning attack as an optimization problem. 

\subsection{Threat Model}
\noindent \textbf{Attacker's Goal.} We consider an attacker's goal is to promote a target item. Specifically, suppose a recommender system recommends top-$K$ items for each user. An attacker's goal is to make its target item appear in the top-$K$ recommendation lists of as many normal users as possible.  We note that an attacker could also aim to demote a target item, making it appear in the top-$K$ recommendation lists of as few normal users as possible. For instance, an attacker may demote its competitor's items. Since demoting a target item can be implemented by promoting other items~\cite{yang2017fake}, we focus on promotion in this work. 

\noindent \textbf{Attacker's Background Knowledge.} We assume an attacker has access to the user-item interaction matrix $\mathbf{Y}$. In many recommender systems such as Amazon and Yelp, users' ratings are public. Therefore, an attacker can write a crawler to collect users' ratings. However, in our experiments, we will also show that our attack is still effective when the attacker has access to a partial user-item interaction matrix. 
The attacker may or may not have access to the internal neural network architecture of the target deep learning based recommender system. 
When the attacker does not have access to the neural network architecture of the target recommender system, the attacker performs attacks by assuming a neural network architecture. 
As we will show in experiments, our attack can transfer between different neural networks, i.e., our attack constructed based on one neural network architecture based recommender system is also effective for other recommender systems that use different neural network architectures.

\noindent \textbf{Attacker's Capabilities.} We assume that an attacker has limited resources, so the attacker can only inject a limited number of fake users. We use $m$ to denote the upper bound of the number of fake users. In addition to the target item, each fake user can rate up to $n$ other items  \blue{to evade trivial detection}. We call these items \textit{filler items}. Specifically, normal users often rate a small number of items, and thus fake users who rate a large number of items are suspicious and can be detected easily. We assume the attacker can inject the fake users' ratings into the training dataset of the target recommender system to manipulate the training process of the deep learning model.

\subsection{Formulating Attacks as an Optimization Problem}\label{sec:opt}

We define the {\bf hit ratio} of an item $t$, denoted as $\mathrm{HR}_t$, as the fraction of normal users who would receive the item $t$ in their top-$K$ recommendation lists. In other words, the hit ratio of $t$ indicates the probability that  $t$ is recommended to a normal user. An attacker's goal is to maximize the hit ratio of a target item $t$. 
Let $\mathbf{y}_{(v)}$ denote the rating score vector of the fake user $v$, and $y_{vi}$ denote the rating score that the fake user $v$ gives to item $i$. A rating score is an element in a set of integers $\{0,1,\ldots,r_{\mathrm{max}}\}$, where $y_{vi}=0$ means that the fake user $v$ has not rated item $i$ and $y_{vi}>0$ represents the preference score fake user $v$ gives to item $i$. For instance, $r_{\mathrm{max}}=5$ in many recommender systems.
Our goal is to craft the ratings for the fake users such that the hit ratio of the target item is maximized. 
Formally, following previous work~\cite{fang2018poisoning}, we formulate crafting the ratings for the fake users as solving the following optimization problem:
\begin{equation}\label{Equ:optobj}
\begin{split}
    \max \quad  & \mathrm{HR}_t\quad \\
    \mathrm{subject\ to} \quad  & ||\mathbf{y}_{(v)}\|_0\leq n+1, \forall v\in \{v_1,v_2,\ldots,v_m\}, \\
    &y_{vi} \in \{0,1,\ldots,r_{\mathrm{max}}\},
\end{split}
\end{equation}
where $\|\mathbf{y}_{(v)}\|_0$ is the number of non-zero entries in fake user $v$'s rating score vector $\mathbf{y}_{(v)}$, 
$n$ is the maximum number of filler items, $m$ is the maximum number of fake users, and $\{v_1,v_2,\ldots,v_m \}$ is the set of $m$ fake users.

\section{Attack Construction: Solving the Optimization Problem}

\begin{figure*}[t]
\centering
\includegraphics[width=0.73\textwidth]{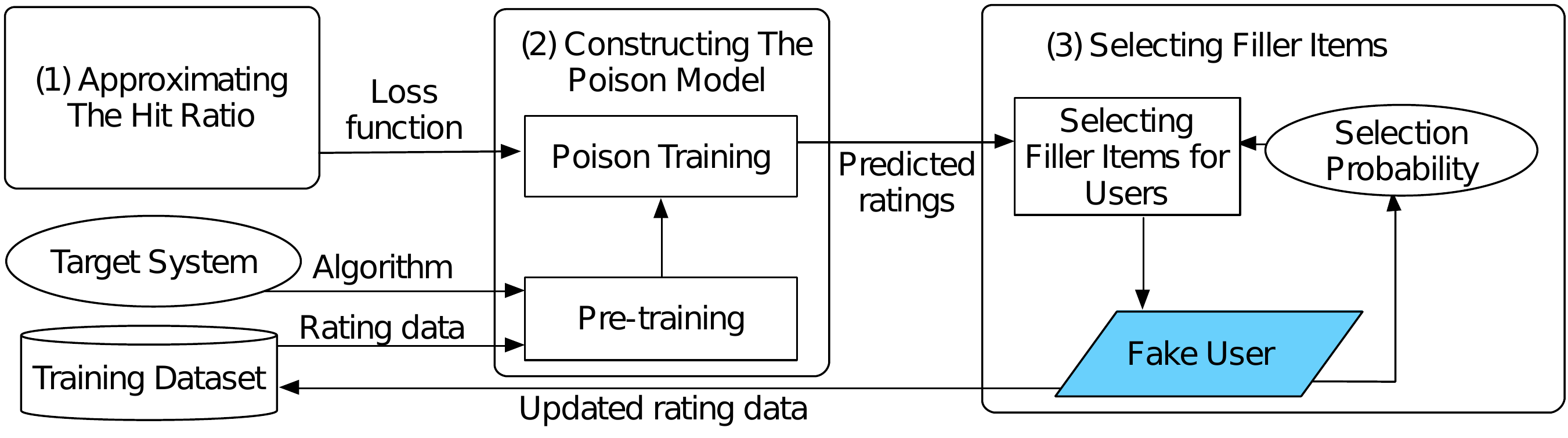}
\caption{An overview of our data poisoning attack. We first use approximation methods to transform the optimization problem into a tractable one and obtain a loss function. Second, according to the obtained loss function, the algorithm used in the target recommender system, and the training dataset, we train a poison model that simulates the compromised target recommender system. Third, we select filler items according to the predicted rankings generated by the poison model and the selection probability. Note that, we will repeat the second and third steps until enough fake users are generated to construct the attack, and the selection probability will be updated in each iteration. }
\label{fig:overview of our attack}
\end{figure*}

\subsection{Overview of Our Proposed Attacks}
\org{A data poisoning attack is essentially to solve the optimization problem in  Eq. (\ref{Equ:optobj}). However, the optimization problem is computationally intractable as it is a non-convex integer programming problem. 
To address the challenge, we develop multiple heuristics to approximately solve the optimization problem. Our heuristics are inspired by previous work~\cite{fang2018poisoning} on attacking graph-based recommender systems.} Figure~\ref{fig:overview of our attack} shows the overview of our data poisoning attack. First,  we approximate the hit ratio using a loss function, where a smaller loss roughly corresponds to a higher hit ratio. Given the loss function, we transform the optimization problem into a tractable one. 
Second, based on our designed loss function, we construct a \textit{poison model} to simulate a compromised deep learning based recommender system. 
In particular, \mage{we first pre-train the poison model to 
ensure that it can correctly predict the preferences of users by using the validation dataset, and then we update the poison model using a loss function, which is derived by extracting the attack related parts in the loss function obtained in the first step,  to approach the expected state of the compromised target recommender system.} 
Third, we select filler items for a fake user according to its  rating score vector predicted  by the poison model and a selection probability vector, where the selection probability vector of items is periodically updated to choose filler items for the next fake user. We repeat the second and third steps until $m$ fake users are generated for the poisoning attack.

\subsection{Approximating the Hit Ratio}\label{sec:approximate}
 The optimization problem we formulated in Eq. (\ref{Equ:optobj}) is computationally intractable because the rating scores are integer variables in the domain $\{0,1,\ldots,r_{\mathrm{max}}\}$ and the hit ratio is a highly non-linear non-differentiable function of the rating scores due to the complexity of the recommender system. To address the computational challenge, we design multiple techniques to convert the optimization problem into a computationally tractable one.

\noindent \textbf{Relaxing Rating Scores to Obtain Continuous Variables.} As for the rating scores in $\mathbf{Y}$ and $\widehat{\mathbf{Y}}$, we can treat them as continuous variables in our attacking process. Specifically, the predicted rating scores, which range from 0.0 to 1.0 in the recommender systems built upon implicit datasets, can be seen as correlations between users and items. After acquiring final rating scores from the target recommender system, we can project them into discrete integer numbers \blue{if necessary}. 

\noindent \textbf{Approximating the Hit Ratio.} The hit ratio $\mathrm{HR}_t$ is the proportion of normal users whose top-$K$ recommendation lists include the target item $t$. Since $\mathrm{HR}_t$ is a highly non-linear non-differentiable function of the users' rating scores, we propose {to} use a loss function to approximate it. {In particular, a smaller loss roughly corresponds to a higher hit ratio.}
Normally, a recommender system uses the \blue{predicted user-item interaction matrix $\widehat{\mathbf{Y}}$} to make recommendations for users. Therefore, we propose to use the following steps to convert the optimization problem as shown in Eq. (\ref{Equ:optobj}).

\begin{enumerate}
\item \textrm{\bf Loss Function for Each User.} We leverage a loss function $l_u$ over the predicted score vector for each user to increase the hit ratio for target item $t$. Intuitively, if the target item $t$ is already included in the recommendation list $L_u$ of user $u$, it is \blue{not necessary} to further improve this recommendation. Otherwise, we should reflect the requirement of the user $u$ in $l_u$ and promote the target item $t$ to get a better ranking among all items. We apply the following loss function for user $u$:
\begin{equation}\label{equ:lu}
    l_u=\max\{\min\limits_{i\in L_u}\log[\widehat{y}_{ui}]-\log[\widehat{y}_{ut}],-\kappa\},
\end{equation}
where $\kappa\geq0$ is a tunable parameter that can be used to enhance the robustness and transferability of our attack. The use of the log operator lessens the dominance effect, while preserving the order of confidence scores due to the monotonicity. As we can see, if a target item $t$ is in $L_u$, $l_u$ will be 0 when $\kappa=0$. Otherwise, larger $\widehat{y}_{ut}$ that is smaller than the minimum value of $\widehat{y}_{ui}$ in $L_u$, larger the positive value of $l_u$ will be. $\kappa$ can ensure the target item $t$ keep a distance from the item with the lowest rating in $L_u$. Thus, we can have a higher probability to include the target item $t$ in the recommendation list of user $u$ by minimizing the loss function $l_u$.

\item \textrm{\bf Loss Function for All Users.} Now we build a loss function for all users. Since our attack goal is to promote the target item to as many users as possible, we design a loss function over all users according to Eq. (\ref{equ:lu}) as follows:
\begin{equation}
    l'=\sum\limits_{u\in S}l_u, 
\end{equation}
where $S$ is the set of all normal users who have not rated target item $t$ yet.

\item \textrm{\bf Converting the Optimization Problem.} 
After relaxing discrete variables to continuous variables and approximating the hit ratio, we can approximate the optimization problem as follows:
\begin{equation}\label{Eq:goal}
    \min G[\mathbf{y}_{(v)}]=\|\mathbf{y}_{(v)}\|_2^2+\eta\cdot l'
\end{equation}
\begin{equation*}
    \mathrm{subject}\ \mathrm{to}\quad y_{vi}\in\lbrack0,r_{\mathrm{max}}\rbrack,
\end{equation*}
where $\eta>0$ is a coefficient to achieve the objective of promoting the target item $t$ \blue{with a limited number of ratings from fake users. Here, we use the $\ell_2$ norm to replace the $\ell_0$ norm in Eq. (\ref{Equ:optobj}), in order to facilitate the calculation of gradients and the stepwise approximation of global optimal values because the $\ell_0$ norm can only compare a limited number of filler item combinations \org{and} cannot continuously change\colorA{, while $\ell_1$ regularization generates sparse rating score vectors, which will reduce  the diversity of the selected filler items for fake users}. 
As for the constraint on the number of filler items, we can achieve it by choosing only a limited number of items for fake user $v$ based on his final rating score vector $\mathbf{y}_{(v)}$.} Thus, we can generate fake users by solving the optimization problem above.
\end{enumerate}

As the users and items in deep learning based recommender systems are completely with discrete labels, gradients will disappear when they  back-propagate to the input layer. Thus, it is infeasible to directly adopt the back-gradient optimization method that has been applied to attack image classifiers ~\cite{munoz2017towards}. 
\colorA{A natural idea is to treat the rating score vector $\mathbf{y}_{(v)}$ of fake user $v$ as the independent variable and formulate the poisoning attack as follows:
\begin{equation}\label{Eq:bi_optim}
    \begin{split}
        \min_{\mathbf{y}_{(v)}}\quad & G[\mathbf{y}_{(v)},\mathbf{w}^{*}]\\
        \mathrm{subject\ to}\quad & \mathbf{w}^{*}=\mathop{\arg\min}_{\mathbf{w}}  \mathcal{L}[\mathbf{w},\mathbf{Y}\cup\mathbf{y}_{(v)}],
    \end{split}
\end{equation}
where $\mathbf{w}^{*}$ represents model parameters, and $\mathcal{L}$ is the original loss function for training the target recommender system. This is a bilevel optimization problem as the lower-level constraint for $\mathbf{w}^{*}$ also depends on $\mathbf{y}_{(v)}$.} 
\colorA{It is quite challenging to solve this optimization problem for deep learning models because the model parameters $\mathbf{w}^{*}$ need to be updated through re-training the model once $\mathbf{y}_{(v)}$ changes. The process would be  time-consuming because it needs to generate enough fake users if we directly compute high order gradients w.r.t. $\mathbf{y}_{(v)}$ and repeat the training process with the whole dataset in each iteration when we gradually update the rating score vector $\mathbf{y}_{(v)}$.}
In particular, we require a large number of of iterations, even thousands of iterations, to accumulate enough changes on the randomly initialized rating score vector for each fake user\colorA{, which is not practical for large recommender systems in the real world}.
Moreover, the \colorA{rating score matrix} used by the recommender systems is usually sparse, and the neural network trained on it might generate predicted rating scores that vary within a certain range, which will be misleading for gradient-based optimization algorithms since they are with small learning rate and can be easily interfered with \colorA{the randomness of model training}.

\subsection{Constructing the Poison Model}\label{sec:poison-model}
Specifically in this step, we construct \org{the} \textit{poison model} to guide the selection of filler items for each fake user according to the obtained loss functions so that we can \blue{efficiently} construct the attack. Here, we investigate and utilize the characteristics of a recommender system itself from a new perspective. For a deep learning based recommender system, as a special type of neural network, it tries to reduce the entropy between users' predicted score vectors and real rating score vectors during the training process. Intuitively, items with higher scores in user $u$'s predicted rating score vector are more likely to have been rated by user $u$ in reality with high scores than other items.  
If we can successfully construct a poison model to simulate
the expected state of the original recommender system after a successful poisoning attack, 
we can infer what kind of fake users in the training dataset can contribute most to the current recommender system. 
The poison model, derived from the initial target recommender system, periodically updates during \blue{the attack} to approach our attack goal gradually. We can then use the poison model to give predictions on fake users' preferences and choose the items with the highest predicted rating scores as filler items for fake users. 

\blue{Note that, the internal structure and hyperparameter settings of the poison model are consistent with the target recommender system.} Moreover, its training datatset should be identical to the original training dataset of the target system initially and can be inserted fake users one by one to simulate attack results. In order to make the poison model change towards our desired objective, we need to define an effective loss function to update the model iteratively. According to the optimization problem in Eq. (\ref{Eq:bi_optim}), we propose the following loss function for the poison model in \blue{the attack}:
\begin{equation}\label{eq:poison_loss}
    l=\colorC{\mathcal{L}}+\lambda\cdot G[\widehat{\mathbf{y}}_{(v)}],
\end{equation}
where \colorC{$\mathcal{L}$} is the loss function chosen in the process of training the original recommender system, e.g., the binary cross entropy over the whole training dataset, $G[\widehat{\mathbf{y}}_{(v)}]$ correlates strongly with our attack goal, and $\lambda>0$ is a coefficient that trades off between the \mage{model validity and the attack objective}, which allows us to generate the poison model close to the recommender system trained under normal circumstances, while achieving our attack goal. 
Here, the validity \colorA{correlates with \colorC{$\mathcal{L}$} and} measures the degree to which the model accurately predicts user preferences on the validation dataset. \colorA{We use the predicted rating score vector $\widehat{\mathbf{y}}_{(v)}$ of fake user $v$ to replace $v$'s real rating score vector $\mathbf{y}_{(v)}$ according to the correlation between them such that we can avoid high order gradient calculation, which is really time-consuming.} 
\blue{Note that, if the \mage{validity} of the poison model is much lower than} \org{that of a model trained normally with the same dataset and the original loss function (i.e., \colorC{$\mathcal{L}$})}\blue{, it is less likely that the poison model approximates the final state of a compromised target recommender system since the target recommender system will always use a validation dataset to guarantee its best performance in the normal model training process. Thus, it is necessary to ensure the \mage{validity} of the poison model during the attack.} In order to make the poison model better simulate the results of the poisoning attack, we design two stages of training: i.e., pre-training and poison training.

\noindent \textbf{Pre-training.} The poison model will be randomly initialized at first and trained on its training dataset with the same loss function (i.e., \colorC{$\mathcal{L}$}) as the target recommender system. After enough iterations, the poison model will be similar to the recommender system obtained from the normal training, ensuring the validity of the model. We can utilize this model to start poison training subsequently.

\noindent \textbf{Poison Training.} The poison model in this stage will use Eq. (\ref{eq:poison_loss}) as loss function and be trained repeatedly w.r.t all model parameters inside it with the back-propagation method. \colorA{We select the initial $\lambda$ such that the loss of the poison model on the validation dataset and the loss that models the attack effectiveness are roughly in the same order of magnitude.} \blue{In the training process, the poison model will get closer to our attack goal and eventually become an ideal state of the target recommender system.} \colorA{We can then use the poison model to help the item selection process for fake users.}

\subsection{Selecting Filler Items}\label{sec:probability}
Now we can select filler items for each fake user based on predicted ratings generated by the acquired final poison model in  the last poison training process.
Note that, items with higher scores in the user's predicted rating score vector given by the recommender system tend to have greater relevance to the user since the system reduces the entropy between users' predicted score vectors and real rating score vectors during the training process. Thus, as long as we get a reasonable poison model, we can obtain the predicted rating score vector $\widehat{\mathbf{y}}_{(v)}$ for fake user $v$ according to the model, and the top-$n$ items \blue{other than target item $t$} will be selected as filler items for fake user $v$.

\blue{However,} as the datasets used in recommender systems are usually very sparse and the models trained from the data \blue{have high randomness}, the recommendation results of deep learning based recommender systems for specific users \blue{and items tend to be} unstable, which means fake users obtained from the poison model may not \colorA{be good choices}. Thus, if we always directly use the predictions of the poison model to select filler items for fake users, we are more likely to \colorA{gradually deviate from the right direction}.  
In order to avoid this, we \blue{develop a concept of \textit{selection probability}, i.e., the probability of an item being selected as filler items.}
We define a selection probability vector as $\mathbf{p}=\{p_1,p_2,\ldots,p_N\}$, each element of which represents the selection probability of the corresponding item. If item $i$ is selected as filler item, $p_i$ will change as follows:
\begin{equation}\label{Eq:p_change}
    p_i=p_i\cdot\delta,
\end{equation}
where $0\leq\delta\leq1$ is an attenuation coefficient that reduces the selection probabilities of selected items. The more times an item is selected as filler item, the lower its selection probability. Note that, $\mathbf{p}$ is initialized to a vector with all element values of 1.0 at first. If all elements in $\mathbf{p}$ are below 1.0 after the  poisoning attack, $\mathbf{p}$ will be initialized again. \blue{After} the poison model gives predicted rating score vector $\widehat{\mathbf{y}}_{(v)}$ for fake user $v$, we combine it with $\mathbf{p}$ to guide filler item selection as follows:
\begin{equation}\label{equ:p}
    \mathbf{r}_v=\widehat{\mathbf{y}}_{(v)}\mathbf{p}^{\mathrm{T}}.
\end{equation}
According to Eq. (\ref{equ:p}), we select these items with the highest $n$ scores in $\mathbf{r}_v$ for fake user $v$ as filler items and update the corresponding selection probabilities \blue{using Eq. (\ref{Eq:p_change})}.
\blue{The use of selection probability refrains from repeated selection of specific items, and provides greater chance of being selected for more candidate items, which allows the target item to build potential correlations with more other items and makes our attack more likely to be effective globally.} As for the recommender systems with sparser datasets, which \blue{means} greater uncertainty of \blue{recommendation results}, we recommend to choose a smaller $\delta$ to strengthen the internal system connectivity, i.e., the target item can correlate more other items, which \blue{avoids} local optimal results and \blue{improves} the attack performance. Combining above insights, we can effectively solve the optimization problem.

The heuristics that solve the optimization problem \mage{are} shown in Algorithm \ref{alg:Framwork}. Note that, since our attack is not designated to specific deep learning recommendation systems and can be generalized to any deep learning based recommender system, the algorithm can opt to solve the problems in various systems. 
\colorA{Our item selection follows three steps. First, we use the poison model to predict a rating score vector $\widehat{\mathbf{y}}_{(v)}$ for a fake user $v$. Second, we compute the element-wise product of $\widehat{\mathbf{y}}_{(v)}$ and a selection probability vector $\mathbf{p}$ as an adjusted rating score vector $\mathbf{r}_v$. Third, we select the $n$ non-target items with the largest adjusted rating scores as the filler items for $v$. For each filler item $i$ for $v$, we decrease its selection probability $p_i$ by multiplying it with a constant (e.g., 0.9) such that it is less likely to be selected as a filler item for other fake users. We use the selection probability vector to increase the diversity of the fake users’ filler items so that the target item can be potentially correlated with more items.} 
Note that we assume the same $n$ is used for each fake user. However, the attacker can also use different number of filler items for different fake users. In particular, an attacker can use our attack to add filler items for a fake user one by one and stop adding filler items once the hit ratio of the target item begins to decrease. 
Finally, we generate rating scores for each filler item according to their previous fitted normal distributions to ensure their scores much similar to other normal ratings, which also be used to effectively evade detection. We will elaborate on the detection performance in Section~\ref{sec:detection}. \colorA{Note that, to speed up the process of generating all fake users, we can also choose to generate $s$ ($s>1$) fake users each time at the cost of reducing the fine-grained control on the attack effectiveness.}

\begin{algorithm}[t]  
  \caption{Our Attack Method}  
  \label{alg:Framwork}  
  \begin{algorithmic}[1]
    \Require  
   User-item interaction matrix $\mathbf{Y}$, target item $t$, parameters $m$, $n$, $K$, $\lambda$, $\eta$, $\kappa$.
    \Ensure  
   $m$ fake users $v_1, v_2, \ldots, v_m$.
   \State // Add fake users one by one
   \For{$v=v_1,v_2, \ldots,v_m$}
   \State Initialize poison model $M_p$ with expanding input user size.
   \State Add the rating tuple $(v, t, r_{\mathrm{max}})$ to $\mathbf{Y}$.
   \State Pre-train $M_p$ on $\mathbf{Y}$ with \colorC{$\mathcal{L}$}. 
   \State Start poison training to get the final poison model $M_p$.
   \State Use $M_p$ to give predicted rating score vector $\widehat{\mathbf{y}}_{(v)}$ for user $v$.
   \State Get $\mathbf{r}_v$ using Eq. (\ref{equ:p}).
   \State Choose these items other than $t$ with the highest $n$ scores in $\mathbf{r}_v$ as filler items.
   \State Update $\mathbf{p}$ using Eq. (\ref{Eq:p_change}).
   \State Generate rating scores for chosen filler items, constituting rating score vector $\mathbf{y}_{(v)}$ for user $v$.
   \State $\mathbf{Y}\gets \mathbf{Y}\cup\mathbf{y}_{(v)}$.
   \EndFor
   \State \textbf{return} $\mathbf{y}_{(v_1)},\mathbf{y}_{(v_2)},\ldots,\mathbf{y}_{(v_m)}$.
  \end{algorithmic}  
\end{algorithm}

\section{Experiments}\label{sec:experiments}
In this section, we first present the \blue{experimental} setup. Second, we evaluate the effectiveness and the transferability of our poisoning attacks. 

\subsection{Experimental Setup}
\noindent \textbf{Datasets.}
We use \colorA{three} real-world datasets to perform our poisoning attacks. They are MovieLens-100K~\cite{harper2015movielens}, \colorA{MovieLens-1M~\cite{harper2015movielens}} and Last.fm~\cite{Cantador:RecSys2011}\colorA{, two different types of typical recommender system datasets}. MovieLens-100K is a classic movie dataset which consists of 943 users, 1,682 movies and 100,000 ratings ranging from 1 to 5. Each user has at least 20 ratings in this dataset. \colorA{Similarly, MovieLens-1M is a larger movie dataset including 6,040 users, 3,706 movies and 1,000,209 ratings ranging from 1 to 5.} Last.fm is a music dataset which contains 1,892 users, 17,632 music artists and 186,479 tag assignments. A user can assign a tag to a music artist, which can be seen as a positive interaction between them. Last.fm is a pure implicit dataset as the tag assignments cannnot be quantified with numerical values. We apply several data processing strategies to make it suitable for our experiments. First, we binarize their interactions as 1.0 for positive ones and 0.0 for others, which can be seen as implicit ratings. Second, we drop the duplicates in the datasets as one user might assign multiple tags to one artist. Third, as the obtained dataset is still sparse, we iteratively filter the dataset to make sure the remaining in the dataset has at least 10 ratings for each user and item (i.e., artist) to avoid the ``cold start'' problem. We end up with a dataset of 701 users, 1,594 items and 36,626 ratings. 

Note that we use implicit training dataset for our target recommender system (i.e., NeuMF), so we also project the ratings in MovieLens-100K \colorA{and MovieLens-1M} to 1.0 when they are larger than 0 and 0.0 otherwise. \colorC{For a binary implicit rating score, 1.0 indicates that the user has rated the item, but it does not necessarily represent that the user likes the item. Likewise, an implicit rating score of 0 does not necessarily represent that the user dislikes the item.}  \colorA{Our attack can be also applied to recommender systems based on explicit ratings since the ratings can be normalized between 0 and 1 before training for such datasets.}

\noindent \textbf{Target Recommender System.}
In our experiments, we evaluate the effectiveness of the attacks by using Neural Matrix Factorization (NeuMF) as the target recommender system. Note that, \colorE{collaborative} filtering systems are one of the most popular and effective recommender systems in the real world. Most websites (e.g., Amazon, YouTube, and Netflix) utilize collaborative filtering as a part of their recommender systems. Moreover, Neural Collaborative Filtering (NCF) is the typical representative of deep learning based recommender systems. Thus, we choose NeuMF, an instantiation of NCF, as the target recommender system since matrix factorization is the most popular technique of collaborative filtering.
    
    

\noindent \textbf{Baseline Attacks.}
We compare our poisoning attack to several existing poisoning attacks. In all these attacks, an attacker injects $m$ fake users to the target recommender system. Different attacks use different strategies to select filler items for fake users. 

\begin{enumerate}
\item {\bf Random Attack}: In a random attack, the attacker randomly chooses $n$ filler items for each fake user. If the training dataset is explicit, the attacker will fit normal distributions on the initial user-item interaction matrix to generate new continuous rating scores for filler items. These rating scores will then be projected to discrete integer numbers if necessary. Even if the training data is implicit, if the initial form of the dataset collected by the recommender system is explicit, the attacker still needs to use the same method to generate rating scores for filler items to evade detection.
\item {\bf Bandwagon Attack}: 
In a bandwagon attack, the popularity of items plays a role in the selection of filler items. We use the average score of the item to represent its popularity on the explicit dataset, and the frequency of the item to represent its popularity on the implicit dataset. We randomly choose $n\times10\%$ items from the set of $10\%$ items with the highest popularity and $n\times90\%$ items among the left unselected items to constitute all filler items. Then we can generate rating scores for filler items with the same method in random attacks.
\item \blue{\bf Poisoning Attack to MF-based Recommender Systems \org{(MF Attack)}}\colorC{~\cite{li2016data,fang2020influence}}: 
\org{MF attack is one of effective poisoning attacks on the most widely used recommender systems, namely matrix-factorization (MF) based recommender systems}.
Note that \org{MF is a traditional and non-deep learning approach for recommender systems, and our work is} the first to implement well-designed poisoning attacks on deep learning based recommender systems, and the target recommender system we conduct experiments on is generally a special kind of matrix-factorization-based recommender systems. 
\colorC{Specifically, we use the PGA attack in \cite{li2016data} as the baseline MF attack in our experiments.} In an MF attack, the attacker will initialize a generalized user-item interaction matrix based on the training dataset at first and then implement optimized poisoning attack on it. We will inject the fake users generated by MF attack into our target deep learning based recommender system to evaluate the effectiveness of this poisoning attack and compare with other poisoning attacks. \colorD{Note that, the existing poisoning attack~\cite{fang2020influence} to matrix-factorization-based recommender systems cannot be directly applied because it requires deriving the influence function for NCF, which can be an interesting topic for future work. Thus, we finally choose the PGA attack in \cite{li2016data} as the MF attack to conduct our experiments.} 
\end{enumerate}

\noindent \textbf{Target Items.}
We evaluate two types of target items in our work, i.e., \textit{random} and \textit{unpopular} target items. Random target items are sampled uniformly at random from the whole item set, while unpopular items are collected randomly from those items with less than \colorA{6 ratings, 10 ratings and 12 ratings for the ML-100K (i.e., MovieLens-100K), ML-1M (i.e., MovieLens-1M), and Music (i.e., Last.fm) datasets, respectively.}  To make our results more convincing, we sample 10 instances for each kind of target items by default and will average their experimental results respectively.

\noindent \textbf{Evaluation Metrics.}
We use the hit ratio of target item $t$ (i.e., $\mathrm{HR}_t@K$) as the metric to evaluate the effectiveness of poisoning attacks for promoting target item $t$. Suppose there are $K$ items in the recommendation list for each user. $\mathrm{HR}_t@K$ is the proportion of normal users whose top-$K$ recommendation lists include target item $t$. We compare $\mathrm{HR}_t@K$ before and after attacks to show the attack effectiveness. As the deep learning based recommender system itself is usually unstable, we train and evaluate the target model on the dataset for 30 times and average the evaluation results. \blue{Note that, for the same kind of target items, we will further use the average value of their hit ratios, recorded as $\mathrm{HR}@K$, to comprehensively evaluate attack performance.}

\noindent \textbf{Parameter Setting.}
Unless otherwise mentioned, the parameter setting for our poisoning attacks is as follows: $\kappa=1$, $\lambda=0.01$, $\eta=100$, and $\delta=0.9$ for the \colorA{ML-100K} dataset; 
\colorA{$\kappa=1$, $\lambda=0.01$, $\eta=100$, and $\delta=0.8$ for the ML-1M dataset};  $\kappa=1$, $\lambda=0.01$, $\eta=100$, and $\delta=0.3$ for the Music dataset; and $m$ equals to \colorA{$1\%$} of the number of normal users, $n=30$, and $K=10$ for \colorA{all} datasets. 
{We conduct our experiments on a CentOS server with 4 NVIDIA  Tesla  V100  GPUs, 64-bit 14-core Intel(R) Xeon(R) CPU E5-2690 v4 @ 2.60GHz and 378 GBs of RAM.} 

\begin{table*}[t]
\centering
  \fontsize{8}{11}\selectfont
  \caption{HR@10 for different attacks with different attack sizes.}
  \label{tab:attack_size_comparison}
    \begin{tabular}{c|c|cccc|cccc}
    \hline
    \multirow{3}{*}{Dataset}&
    \multirow{3}{*}{Attack}&
    \multicolumn{8}{c}{Attack size}\cr\cline{3-10}
    &&\multicolumn{4}{c|}{Random target items}&\multicolumn{4}{c}{Unpopular target items}\cr\cline{3-10}
    &&0.5\%&1\%&3\%&5\%&0.5\%&1\%&3\%&5\%\cr
    \hline
    \hline
    \multirow{5}{*}{ML-100K}&None&0.0025&0.0025&0.0025&0.0025&0&0&0&0\cr
    &Random&0.0028&0.0034&0.0053&0.0078&0.0002&0.0003&0.0013&0.0025\cr
    &Bandwagon&0.0030&0.0034&0.0055&0.0081&0.0002&0.0004&0.0013&0.0024\cr
    &MF&0.0032&0.0035&0.0069&0.0090&0.0001&0.0002&0.0014&0.0033\cr
    &Our attack&{\bf 0.0034}&{\bf 0.0046}&{\bf 0.0100}&{\bf 0.0151}&{\bf 0.0007}&{\bf 0.0019}&{\bf 0.0111}&{\bf 0.0206}\cr
    \hline
    \multirow{5}{*}{Music}&None&0.0024&0.0024&0.0024&0.0024&0.0003&0.0003&0.0003&0.0003\cr
    &Random&0.0037&0.0048&0.0115&0.0216&0.0006&0.0014&0.0053&0.0118\cr
    &Bandwagon&0.0036&0.0046&0.0104&0.0176&0.0005& 0.0011&0.0044&0.0094\cr
    &MF&0.0034&0.0050&0.0120&0.0210&0.0005&0.0017&0.0058&0.0118\cr
    &Our attack&{\bf 0.0047}&{\bf 0.0068}&{\bf 0.0144}&{\bf 0.0243}&{\bf 0.0012}&{\bf 0.0026}&{\bf 0.0086}&{\bf 0.0161}\cr
    \hline
    \hline
    \end{tabular}

\end{table*}

\subsection{Effectiveness of Poisoning Attacks}\label{sec:effect_poison}

Now we conduct our experiments under the white-box setting. Under this setting, we assume that \blue{the attacker is aware of} the internal structure, the training data and the hyperparameters of the target recommender system so that we can train an initial poison model that has similar functions with the target recommender system.

\noindent \textbf{Impact of the Number of Inserted Fake Users.} 
Table \ref{tab:attack_size_comparison} shows the results of poisoning attacks with different number of fake users. 
We measure the effectiveness of attacks with different  attack sizes, i.e., the fraction of the number of fake users to that of original normal users. In the table, ``None'' means no poisoning attacks performed on the target recommender system and \blue{MF represents the poisoning attack method on matrix-factorization-based recommender systems.} 
We find that our attack is very effective in promoting target items on both datasets. For example, after inserting only \colorA{$5\%$ fake users into the Music dataset, the hit ratio for random target items increases by about \colorB{9.1} times.} 

Also, we observe that the attack performance of all attack methods increases as the number of fake users inserted increases. For instance, after injecting \colorA{$0.5\%$ fake users for random target items to the \colorA{ML-100K} dataset, our attack can achieve a hit ratio of 0.0034, while the hit ratio increases to 0.0151 when injecting $5\%$ fake users.} The results are reasonable because, when more fake users are inserted, the target items occur more in the poisoned training dataset and thus can influence the recommender system more significantly. 

Our attack significantly outperforms the \colorB{baseline} attacks in all cases. As for the \colorA{ML-100K} dataset, our attack is quite outstanding and comprehensively surpasses all compared methods. In particular, when inserting \colorA{$5\%$ fake users for unpopular target items, our attack achieves the hit ratio of 0.0206, about 6.2 times of the best hit ratio obtained by other attacks}. With the Music dataset, our attack is still the most effective for all situations. For instance, our attacks can improve the hit ratio of unpopular target items from 0.0003 to \colorA{0.0086 with an attack size of $3\%$}. The MF attack achieves the best performance among the baseline attacks. It can increase the hit ratio to 0.0058, which is only $67.4\%$ of that of our attack. The possible reason is that the random attack and the \colorB{bandwagon} attack do not leverage the information of deep learning models, e.g., the model structure and parameters, so that they cannot perform well on deep learning based systems. The MF attack is designed for factorization-based recommender systems that use linear inner product of latent vectors to recover $\mathbf{Y}$, while the target deep learning based recommender system in our experiments uses extra nonlinear structure. Thus, the MF attack cannot achieve good attack effectiveness as our attack. 

\colorA{To further evaluate the effectiveness of poisoning attacks on large datasets, we conduct the experiments on the ML-1M dataset with an attack size of 5\% and sample \colorC{5} items for each type of target items. Note that, to speed up our poisoning attack, we generate 5 fake users each time. The results are shown in Table \ref{tab:large_scale}. First, we observe that, similar to the small datasets, the large dataset is also vulnerable to poisoning attacks. The hit ratio of unpopular target items increases from 0 to 0.0034 and 0.0060 with the bandwagon attack and our attack, respectively. Second, our attack still performs the best among all poisoning attacks on both random target items and unpopular ones. For example, the hit ratio of random target items under our attack is 0.0099, about 1.2 times of the highest result among the baseline attacks.}

Moreover, the increase of the hit ratio of unpopular target items is much more significant than that of random target items. For instance, when injecting $5\%$ fake users into the Music dataset, the hit ratio of our attacks for random target items increases by around \colorB{9.1} times compared with initial hit ratio while that of unpopular items increases by about \colorB{52.7} times. We suppose that it is caused by the existence of competing items. When the hit ratio of the target item increases, the hit ratios of other items correlated with it also tend to increase. As the sum of all hit ratios is fixed (i.e., 1), there will be a competitive relationship between them when the hit ratios on both sides rise to a large value. Unpopular items have few correlated items since they have few ratings in the original dataset. Therefore, after a successful attack, there will be fewer items competing with them than random target items. This result is encouraging because the items that attackers want to promote are usually unpopular ones. 

Furthermore, all poisoning attacks on the Music dataset are more effective than the \colorA{ML-100K and ML-1M datasets}. For example, when promoting random target items with our attack method, an attacker can increase the hit ratio by about \colorB{5.0 times, 4.8 times, and 9.1 times for the ML-100K, ML-1M and Music dataset, respectively, after injecting $5\%$ fake users.} The possible reason is that the Music dataset is more sparse, making the recommender system less stable and more vulnerable to the poisoning attacks.
\colorC{
The standard deviations of the hit ratios in Table \ref{tab:attack_size_comparison} can be found in Appendix \ref{sec:standard_diviations}, and the change of the hit ratio for each target item is presented in Appendix \ref{sec:HR_per_target}. These results further demonstrate the effectiveness of our attack.}

\begin{table}[t]
\centering
  \fontsize{8}{11}\selectfont
  \caption{HR@10 on a large dataset.}
  \label{tab:large_scale}
    \begin{tabular}{c|c|c|c}
    \hline
    \multirow{2}{*}{Dataset}&\multirow{2}{*}{Attack}&
    \multicolumn{2}{c}{Target items}\cr\cline{3-4}
    &&Random&Unpopular\cr
    \hline
    \hline
    \multirow{5}{*}{ML-1M}&None&0.0017&0\cr
    &Random&0.0069&0.0024\cr
    &Bandwagon&0.0080&0.0034\cr
    &MF&0.0060&0.0029\cr
    &Our attack&{\bf 0.0099}&{\bf 0.0060}\cr
    \hline
    \hline
    \end{tabular}

\end{table}

\noindent \textbf{Impact of the Number of Recommended Items. } 
Table \ref{tab:topk_comparison} shows the results of  poisoning attacks with different numbers of recommended items (i.e., $K$) in a recommendation list. Attack size for all poisoning attacks is set to $1\%$ and the number of filler items (i.e., $n$) is set to by default 30 for all methods. We choose unpopular target items to conduct our experiments. First, we observe that our attack is still the most effective method among all the poisoning attacks in all cases, e.g., when $K=20$, the hit ratio of our attack on the \colorA{ML-100K dataset is about 6.0 times of the best hit ratio achieved by the baseline attacks}. On the Music dataset, we can observe similar results. For example, the MF attack can increase the hit ratio to 0.0040 when $K=20$,  which is the best among the existing methods, while our attack achieves the performance of 0.0061, about 1.5 times of the former.

The hit ratios of all methods tend to increase with $K$. As we can see, the initial hit ratio with no injected fake users increase when $K$ increases on the Music dataset. Similarly, hit ratios for all poisoning attacks gradually become larger when $K$ increases on \colorB{both} datasets. For instance, the hit ratio of our attack on the \colorA{ML-100K dataset when $K=20$ is about 3.5 times of that when $K=5$}. A larger $K$ means a greater chance for target items to be included in the recommendation list. This phenomenon is particularly obvious and significant in our attack. 

\begin{table}[t]
\centering
  \fontsize{8}{11}\selectfont
  \caption{HR@$K$ for different $K$.}
  \label{tab:topk_comparison}
    \begin{tabular}{c|c|cccc}
    \hline
    \multirow{2}{*}{Dataset}&
    \multirow{2}{*}{Attack}&
    \multicolumn{4}{c}{$K$}\cr\cline{3-6}
    &&5&10&15&20\cr
    \hline
    \hline
    \multirow{5}{*}{ML-100K}&None&0&0&0&0\cr
    &Random&0.0002&0.0003&0.0005&0.0006\cr
    &Bandwagon&0.0002&0.0004&0.0006&0.0007\cr
    &MF&0.0002&0.0002&0.0004&0.0006\cr
    &Our attack&{\bf 0.0012}&{\bf 0.0019}&{\bf 0.0033}&{\bf 0.0042}\cr
    \hline
    \multirow{5}{*}{Music}&None&0.0001&0.0003&0.0005&0.0007\cr
    &Random&0.0005&0.0014&0.0025&0.0037\cr
    &Bandwagon&0.0003&0.0011&0.0018&0.0027\cr
    &MF&0.0006&0.0017&0.0029&0.0040\cr
    &Our attack&{\bf 0.0007}&{\bf 0.0026}&{\bf 0.0042}&{\bf 0.0061}\cr
    \hline
    \hline
    \end{tabular}

\end{table}

\begin{figure*}[!t]
\centering
\subfloat[ML-100K]{\includegraphics[width=3in]{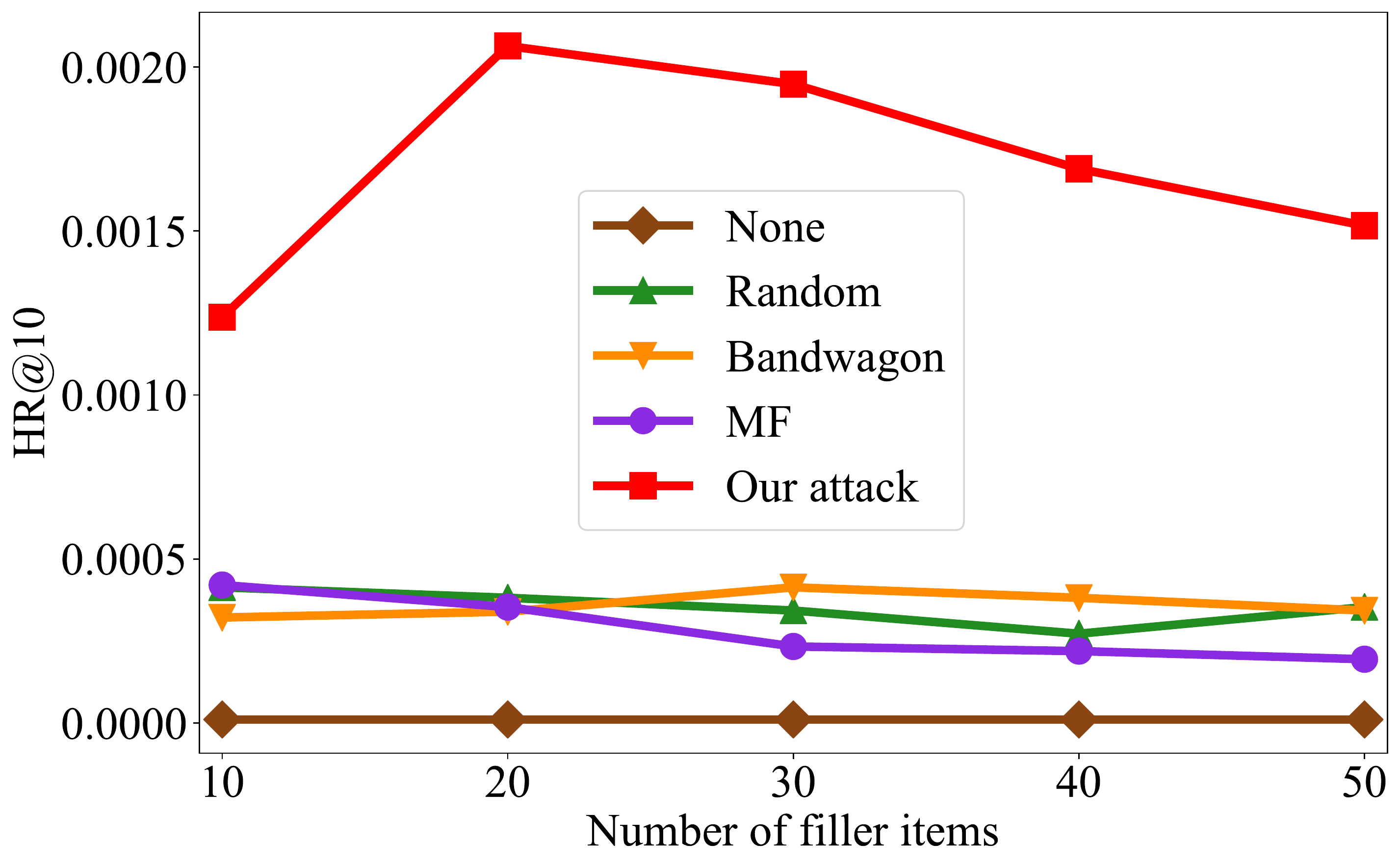}
\label{ml-100k_diff_n}}
\hfil
\subfloat[Music]{\includegraphics[width=3in]{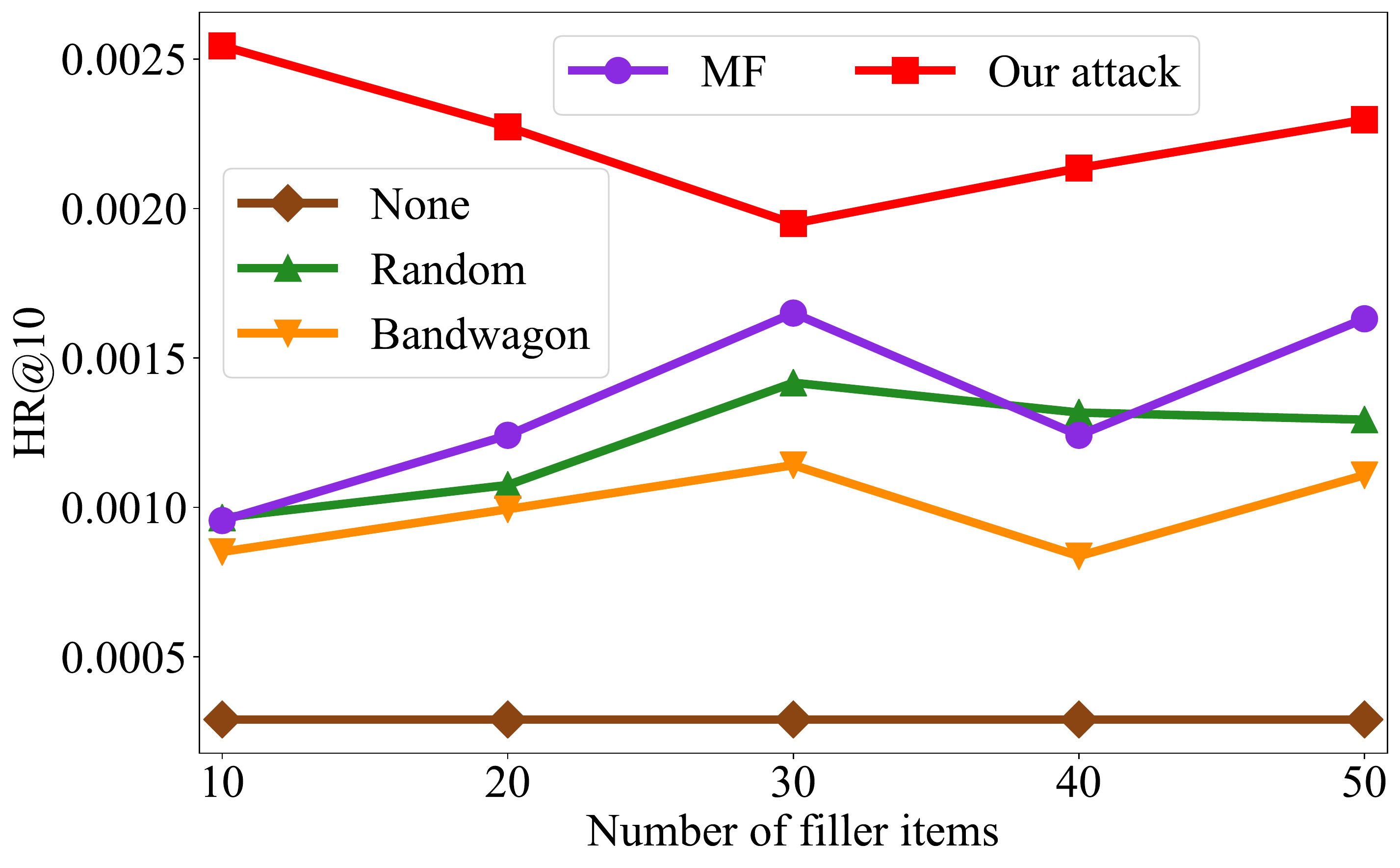}
\label{lastfm_diff_n}}
\caption{The impact of the number of filler items on the attack effectiveness.}
\label{fig_sim}
\end{figure*}

\noindent \textbf{Impact of the Number of Filler Items. } 
Figure \ref{fig_sim} illustrates the results of poisoning attacks with different numbers of filler items (i.e., $n$) that changes from 10 to 50. We choose unpopular target items as our target items. We have some interesting observations. First, our attack always outperforms other existing poisoning attacks in all test cases, which further demonstrates the effectiveness and robustness of our attack. In particular, when $n$ is relatively small, the performance of our attack is still the best. Therefore, when an attacker tries to insert as few filler items as possible to evade detection by the target recommender system, our attack method is the best \mage{choice} to implement effective attacks.
Second, the hit ratio may not always increase when $n$ increases. On the ML-100K dataset, the performance of our attack increases first and then dicreases with the increase of $n$. It achieves the best result when $n=20$. The attack effectiveness of the MF attack tends to decrease when $n$ increases, while other attacks achieve relatively stable performance. However, on the Music dataset, the hit ratio of our attack descends first and then ascends with the increase of $n$, while the hit ratios of other attacks fluctuate. These results show that there is no linear correlation between the attack effectiveness and $n$. As with different datasets, the most suitable $n$ can be different for the existing poisoning attacks. We suppose that, when $n$ is small, each fake user can only exert limited influence on the target recommender system, while, when $n$ is large, there might be some items that are ineffective in promoting the target item, and even competing items included in filler items. Thus, the best number of filler items is closely related to the attack methods and the used datasets.

\begin{figure*}[!t]
\centering
\subfloat[ML-100K]{\includegraphics[width=3in]{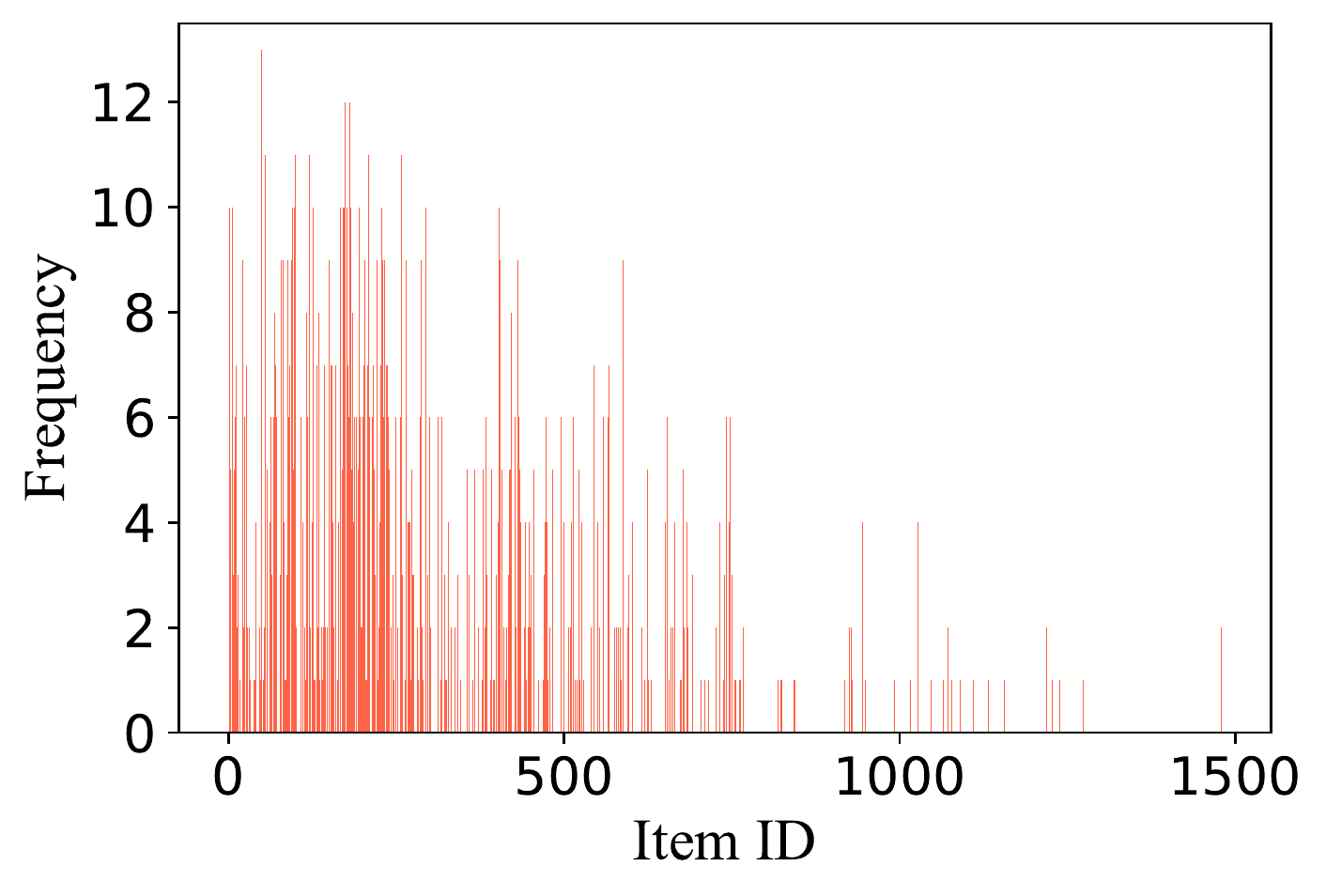}
\label{ml-100k_diversity}}
 \hfil
\subfloat[Music]{\includegraphics[width=2.9in]{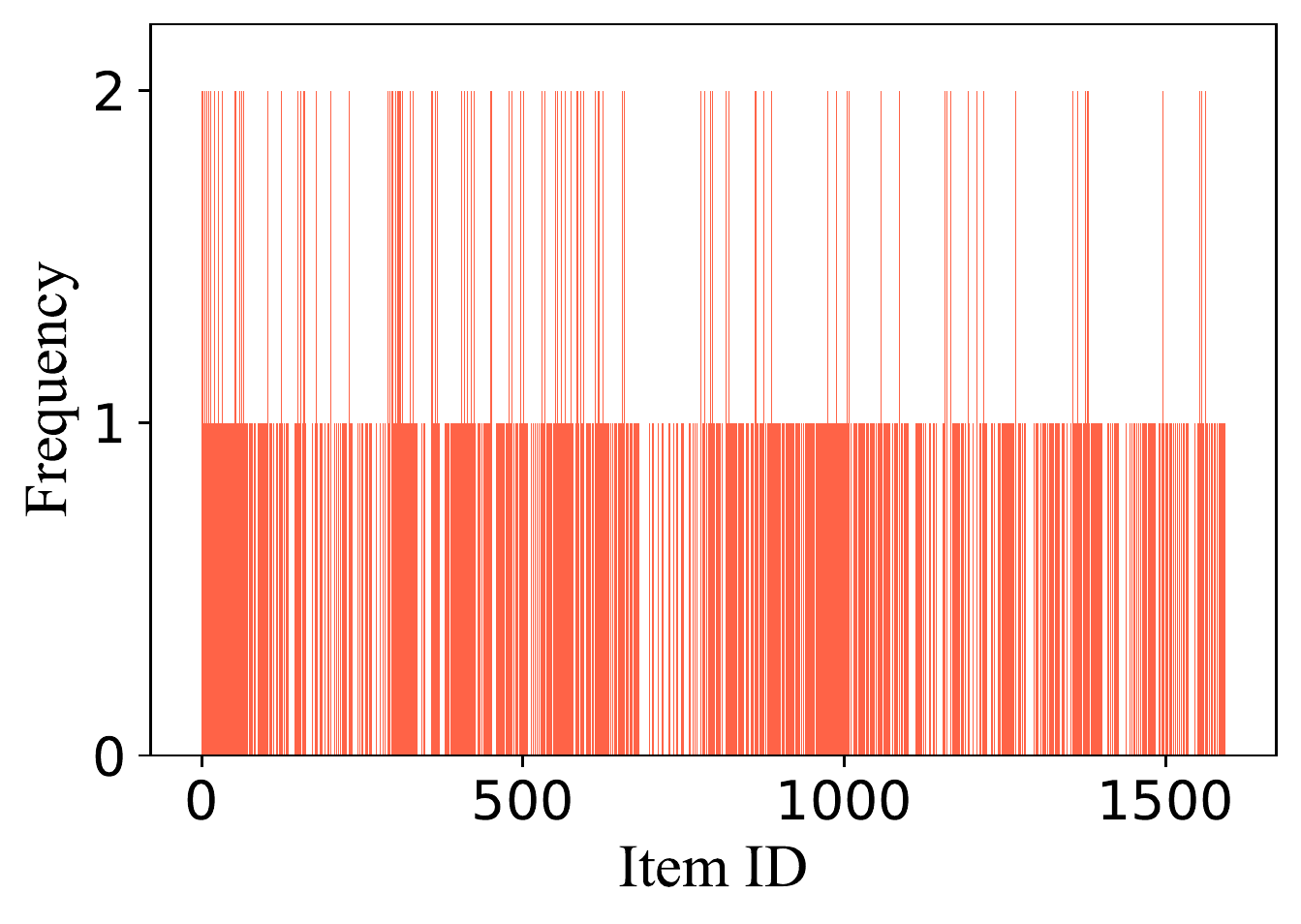}
\label{lastfm_diversity}}
\caption{The impact of $\delta$ on the diversity of filler items.}
\label{fig_diversity}
\end{figure*}

\noindent \colorC{\textbf{Impact of $\delta$.} As an important parameter used in our attack, $\delta$ can affect the diversity of filler items and further impact the attack effectiveness. We select two random target items from the ML-100K dataset and the Music dataset respectively and analyze the diversity of the filler items selected by our attack. For simplicity, we inject $5\%$ fake users. The results are shown in Figure \ref{fig_diversity}. First, we observe that filler items on both datasets have good diversities. The highest frequency of filler items on the ML-100K dataset is 13, around $1.4\%$ of the total number of normal users, and all other items have relatively low frequency. On the Music dataset, the frequency of all filler items is not larger than 2, indicating a strong diversity. Second, the filler items on the Music dataset have a stronger diversity than that on the ML-100K dataset. The filler items on the Music dataset are more evenly distributed than those on the ML-100K dataset and the average of their frequency is lower than that of the ML-100K. The reason is that we use a smaller $\delta$ for the Music dataset, which ensures a better diversity.}

\begin{figure}[!t]
\centering
\includegraphics[width=0.43\textwidth]{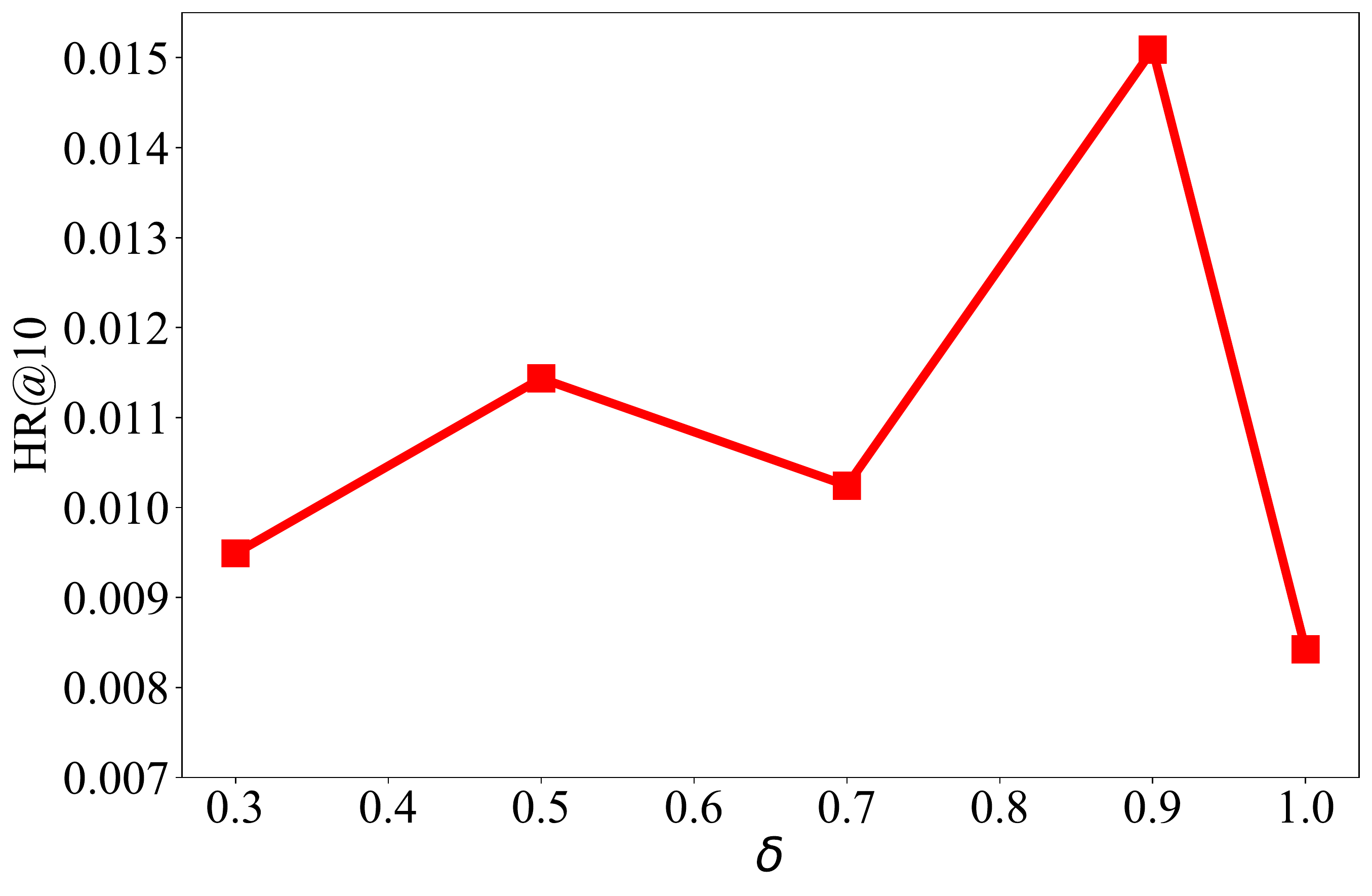}
\caption{The impact of $\delta$ on the attack effectiveness for the ML-100K dataset. }
\label{fig:delta}
\end{figure}

\colorC{To further investigate the impact of $\delta$ on the attack effectiveness, we change the value of $\delta$ and inject $5\%$ fake users on the ML-100K dataset for random target items. The results are illustrated in Figure \ref{fig:delta}. First, we observe that $\delta$ has a significant influence on the attack effectiveness of our method on the ML-100K dataset. \colorD{The hit ratio for target items does not always ascend when $\delta$ increases, and the best $\delta$ for the ML-100K dataset is around 0.9.} Second, compared to the hit ratio of target items when $\delta=1$, i.e., no change is required for the selection probability vector after generating a fake user, $\delta$ helps to promote the attack effectiveness when $\delta<1$. Third, our attack still outperforms other attack methods in most cases, which demonstrates the robustness of our attack.}

\noindent \colorA{\textbf{Impact of the Target Item Rated by Fake Users.} We assume that each fake user will certainly rate the target item in the attacks including our attack as well as the baseline attacks. It is inspired by the observation that the most effective method to promote an item is to assign it with high rating scores in the training dataset due to the strong correlation between users' real ratings and predicted ratings. To accurately evaluate the impact of the target item rated by fake users, we now consider what if the target item is not rated by fake users by default in various poisoning attacks. Note that, we set $\delta=1.0$ here and select the $(n+1)$ items with the highest adjusted predicted rating scores as those items rated by the fake user in our attack\colorB{, and the baseline attacks follow their own rules to select the ($n+1$) rated items for fake users}. We choose the random target items in the ML-100K dataset to conduct the experiments. The experimental results are shown in Table \ref{tab:no_target}. Compared to the results presented in Table \ref{tab:attack_size_comparison}, we can observe that the effectiveness of all attack methods is reduced significantly when the target items are not selected by default. However, our method remains effective as the hit ratio of target items still increases by 1.6 times when injecting $5\%$ fake users, while other baseline attacks are ineffective in this scenario.}

\begin{table}[t]
\centering
  \fontsize{8}{11}\selectfont
  \caption{HR@10 for different attacks without target items selected by default.}
  \label{tab:no_target}
    \begin{tabular}{c|c|cccc}
    \hline
    \multirow{2}{*}{Dataset}&
    \multirow{2}{*}{Attack}&
    \multicolumn{4}{c}{Attack size}\cr\cline{3-6}
    &&0.5\%&1\%&3\%&5\%\cr
    \hline
    \hline
    \multirow{5}{*}{ML-100K}&None&0.0025&0.0025&0.0025&0.0025\cr
    &Random&0.0025&0.0025&0.0024&0.0025\cr
    &Bandwagon&0.0026&0.0026&0.0028&0.0024\cr
    &MF&0.0026&0.0026&0.0027&0.0025\cr
    &Our attack&{\bf 0.0028}&{\bf 0.0034}&{\bf 0.0043}&{\bf 0.0064}\cr
    \hline
    \hline
    \end{tabular}
\end{table}

\begin{table*}[tp]
\centering
  \fontsize{8}{11}\selectfont
  \caption{HR@10 under \colorC{the transferability} setting.}
  \label{tab:blackbox_comparison}
    \begin{tabular}{c|c|cccc|cccc}
    \hline
    \multirow{3}{*}{Dataset}&
    \multirow{3}{*}{Attack}&
    \multicolumn{8}{c}{Attack size}\cr\cline{3-10}
    &&\multicolumn{4}{c|}{Random target items}&\multicolumn{4}{c}{Unpopular target items}\cr\cline{3-10}
    &&0.5\%&1\%&3\%&5\%&0.5\%&1\%&3\%&5\%\cr
    \hline
    \hline
    \multirow{5}{*}{ML-100K}&None&0.0023&0.0023&0.0023&0.0023&0&0&0&0\cr
    &Random&0.0027&0.0035&0.0070&0.0083&0.0002&0.0005&0.0016&0.0030\cr
    &Bandwagon&0.0027&0.0030&0.0070&0.0092&0.0003&0.0005&0.0018&0.0034\cr
    &MF&0.0027&0.0036&0.0064&0.0096&0.0003&0.0005&0.0019&0.0035\cr
    &Our attack&{\bf 0.0038}&{\bf 0.0042}&{\bf 0.0099}&{\bf 0.0150}&{\bf 0.0010}&{\bf 0.0023}&{\bf 0.0082}&{\bf 0.0141}\cr
    \hline
    \multirow{5}{*}{Music}&None&0.0009&0.0009&0.0009&0.0009&0.0001&0.0001&0.0001&0.0001\cr
    &Random&{\bf 0.0020}&0.0024&0.0088&0.0189&0.0003& 0.0010&0.0042&0.0101\cr
    &Bandwagon&0.0011&0.0025&0.0074&0.0160&0.0001&0.0004&0.0027&0.0091\cr
    &MF&0.0015&{\bf 0.0028}&0.0087&0.0152&0.0004&0.0009&0.0049&0.0096\cr
    &Our attack&0.0015&0.0022&{\bf 0.0128}&{\bf 0.0214}&{\bf 0.0007}& {\bf 0.0014}&{\bf 0.0101}&{\bf 0.0184}\cr
    \hline
    \hline
    \end{tabular}
\end{table*}

\subsection{Attacks with Partial Knowledge}\label{subsec:partial}

\colorC{In the experiments above, we assume that an attacker has full access to the whole dataset of the target recommender system, which does not always hold in practice. The attacker may only have partial knowledge of the dataset. To evaluate the effectiveness of different poisoning attacks under this setting, we conduct further experiments with two different types of partial knowledge. One partial knowledge is that the attacker knows partial rating scores of all normal users, and the other is that the attacker knows all rating scores of only partial normal users. Note that, all these experiments are evaluated on the original dataset that contains all users and all rating scores. \colorD{We use the random target items in our experiments.}
The results are shown in Table \ref{tab:partial_matrix} and Table \ref{tab:partial_users}, respectively. According to Table \ref{tab:partial_matrix}, we observe that, even with only $30\%$ ratings of the original rating matrix, the hit ratio of the random target items in our attack is 0.0092, which is only slightly smaller than that with full knowledge, i.e., 0.0099, (see Table \ref{tab:large_scale}) and much larger than the best result of baseline attacks, i.e., 0.0069 achieved by the random attack. However, the bandwagon attack and the MF attack are much less effective with only partial knowledge. Similarly, in Table \ref{tab:partial_users}, our attack still outperforms the  baseline attacks, and the hit ratio of our attack is only slightly smaller than that with full knowledge. The results demonstrate that our attack is still effective even when the attacker only has partial knowledge of the training data, while the bandwagon attack and the MF attack heavily relies on the information informed from the observed dataset.}

\begin{table}[t]
\centering
  \fontsize{7.5}{11}\selectfont
  \caption{HR@10 on ML-1M dataset with a partial rating matrix.}
  \label{tab:partial_matrix}
    \begin{tabular}{c|c|c}
    \hline
    \multirow{1}{*}{Knowledge level}&\multirow{1}{*}{Attack}&Random target items\cr
    \hline
    \hline
    \multirow{5}{*}{30\%}&None&0.0017\cr
    &Random&0.0069\cr
    &Bandwagon&0.0060\cr
    &MF&0.0040\cr
    &Our attack&{\bf 0.0092}\cr
    \hline
    \hline
    \end{tabular}
    
\end{table}

\begin{table}[t]
\centering
  \fontsize{7.5}{11}\selectfont
  \caption{HR@10 on ML-1M dataset with a subset of users.}
  \label{tab:partial_users}
    \begin{tabular}{c|c|c}
    \hline
    \multirow{1}{*}{Knowledge level}&\multirow{1}{*}{Attack}&Random target items\cr
    \hline
    \hline
    \multirow{5}{*}{30\%}&None&0.0017\cr
    &Random&0.0069\cr
    &Bandwagon&0.0057\cr
    &MF&0.0035\cr
    &Our attack&{\bf 0.0091}\cr
    \hline
    \hline
    \end{tabular}
    
\end{table}

\subsection{Transferability}
In the previous experiments, we assume a white-box setting under which an attacker knows the internal structure, the training data and the hyperparameters of the target recommender system. As long as we use the known data and the model structure to train a surrogate model locally, we can obtain a model having a similar function to the target recommender system. \colorC{To further evaluate the transferability of our attack, we consider the gray-box setting under which the attacker only knows the algorithm and the training data used by the target recommender system.}

We assume that an attacker generates fake users based on a surrogate model that is different from the internal structure of the target recommender system. Specifically, we change the number of MLP layers to constitute a different target recommender system. Note that, these target items and filler items generated for all fake users under this setting are consistent with that under the white-box setting. Table \ref{tab:blackbox_comparison} shows the hit ratios of our attacks and the existing attacks for both random and unpopular target items on two datasets.
First, both our attack and the existing attacks can \blue{ increase the hit ratio of target items notably}. \colorA{For instance, our method increases the hit ratio of random target items by about \colorB{22.8} times compared to the initial hit ratio when the attack size is $5\%$ on the Music dataset.}

Second, our method shows the best transferring \blue{effectiveness} in \blue{most} situations, which means that our method has better transferability than the existing attacks. For example, our attack achieves the highest hit ratio of random target items, \colorA{i.e., 0.0150, with an attack size of $5\%$ on the ML-100K dataset, which is about 1.6 times} of the best \blue{performance of the} baseline attacks. Similarly, on the Music dataset, our attack increases the hit ratio of unpopular target items from \colorA{0.0001 to 0.0184 by injecting $5\%$ fake users, while the existing attacks obtain the highest hit ratio of 0.0101, which is $54.9\%$ of ours.} 

Third, similar to the results under the white-box setting, we can observe that the increase of the hit ratio on the Music dataset is more notable than \colorA{\colorB{that} on the ML-100K dataset}. For instance, with an attack size of \colorA{$5\%$ on random target items, our attacks can increase the hit ratio by around \colorB{22.8 times and 5.5 times} on the Music and the ML-100K datasets, respectively, compared with the corresponding initial hit ratios}. The reason is that the Music dataset is more sparse, which makes the recommender systems trained on it less stable and easier to be compromised.

\colorC{In summary, our attack achieves a better transferability than the baseline attacks, which means that our attack poses a greater threat to unknown target recommender systems.}

\section{Detecting Fake Users}~\label{sec:detection}

In this section, we evaluate the effectiveness of the attack under a detector built upon \blue{rating} scores. Detecting fake users is also known as \textit{Sybil detection}. Many methods have been proposed for Sybil detection. These methods leverage user registration information~\cite{yuan2019detecting}, user-generated content~\cite{wang2013you,cao2014uncovering}, and/or social graphs between users~\cite{yu2006sybilguard,danezis2009sybilinfer,gong2014sybilbelief,jia2017random,wang2017gang,wang2017sybilscar,wang19ndss}. 
Since we have no access to users' registration information and social graphs, similar to~\cite{fang2018poisoning}, we utilize a detection method based on user-generated content, i.e., the ratings of users on items. We extract useful features from the datasets and generate certain feature values for each user. We train a fake user classifier \blue{for each poisoning attack} to detect fake users. 
We will study the effectiveness of the poisoning attacks when the recommender system has deployed such a detector.  

\noindent \textbf{Rating Score Based Detection.} Similar to the existing defenses~\cite{chirita2005preventing,mobasher2007toward,fang2018poisoning} that \blue{leverage} several statistical features from rating scores to distinguish normal users from fake users, we adopt these features to train our detection classifiers. The details of these features are described as follows.

\begin{itemize}
\item \textit{Rating Deviation from Mean Agreement (RDMA)} ~\cite{chirita2005preventing}. The feature indicates the average deviation of rating scores of a user to the mean rating scores of the corresponding items, which is computed as follows for a user $u$:
\begin{equation}
    \mathrm{RDMA}_u=\frac{\sum\limits_{i\in I_u}\frac{\mid y_{ui}-\overline{\mathbf{y}^{(i)}}\mid}{c_i}}{\mid I_u\mid},
\end{equation}
where $I_u$ is the set of items that user $u$ has rated, $\mid I_u\mid$ is the number of items in $I_u$ , $y_{ui}$ is user $u$'s  ratings score for item $i$, $\overline{\mathbf{y}^{(i)}}$ is the average rating score of item $i$, and $c_i$ is
the total number of ratings for item $i$ in the whole dataset.

\item \textit{Weighted Degree of Agreement (WDA)}~ \cite{mobasher2007toward}. The feature is the numerator of the RDMA feature, which is computed as follows:
\begin{equation}
    \mathrm{WDA}_u=\sum_{i\in I_u}\frac{\mid y_{ui}-\overline{\mathbf{y}^{(i)}}\mid}{c_i}.
\end{equation}

\item \textit{Weighted Deviation from Mean Agreement (WDMA)}~\cite{mobasher2007toward}.  
This feature considers more the items that have less ratings, which is similar in form to RDMA. It is calculated as follows:
\begin{equation}
    \mathrm{WDMA}_u=\frac{\sum\limits_{i\in I_u}\frac{\mid y_{ui}-\overline{\mathbf{y}^{(i)}}\mid}{c_i^2}}{\mid I_u\mid}.
\end{equation}

\item \textit{Mean Variance (MeanVar)} \cite{mobasher2007toward}. This feature denotes the average variance of rating scores of a uesr to the mean rating scores of the corresponding items. The MeanVar feature for a user $u$ is computed as follows:
\begin{equation}
    \mathrm{MeanVar}_u=\frac{\sum\limits_{i\in I_u}[y_{ui}-\overline{\mathbf{y}^{(i)}}]^2}{\mid I_u\mid}.
\end{equation}

\item \textit{Filler Mean Target Difference (FMTD)}~\cite{mobasher2007toward}. This feature measures the divergence between rating scores of a user, which is obtained by:
\begin{equation}
    \mathrm{FMTD}_u=\left|\frac{\sum\limits_{i\in I_{uM}}y_{ui}}{\mid I_{uM}\mid}-\frac{\sum\limits_{j\in I_{uO}}y_{uj}}{\mid I_{uO}\mid}\right|,
\end{equation}
where $I_{uM}$ is the set of items in $I_u$ that $u$ gave the maximum rating score and $I_{uO}$ includes all other items in $I_u$.
\end{itemize}

\begin{table*}[t]
\centering
  \fontsize{8}{11}\selectfont
  \caption{Detection results for different attacks.}
  \label{tab:detection_results}
    \begin{tabular}{c|c|c|cccc|cccc}
    \hline
    \multirow{2}{*}{Dataset}&
    \multirow{2}{*}{Phase}&
    \multirow{2}{*}{Attack}&
    \multicolumn{4}{c|}{FPR}&\multicolumn{4}{c}{FNR}\cr\cline{4-11}
    &&&0.5\%&1\%&3\%&5\%&0.5\%&1\%&3\%&5\%\cr
    \hline
    \hline
    \multirow{8}{*}{ML-100K}&
    \multirow{4}{*}{SVM}&Random&0.0106&0.0106&0.0106&0.0106&0.0200&0.0111&0.0179&0.0021\cr
    &&Bandwagon&0.0127&0.0127&0.0127&0.0127&0.0400&0.0222&0.0214&0.0127\cr
    &&MF&0.0191&0.0191&0.0191&0.0191&0.0400&0.0556&0.0500&0.0298\cr
    &&Our attack&{\bf 0.1410}&{\bf0.1410}&{\bf 0.1410}&{\bf 0.1410}&{\bf 0.3400}&{\bf 0.3444}&{\bf 0.2357}&{\bf 0.2340}\cr
    \cline{2-11}&
    \multirow{4}{*}{TIA}&Random&0.0001&0.0001&0.0001&0.0001&0.0200&0.0111&0.0179&0.0021\cr
    &&Bandwagon&0&0&0.0003&0.0008&0.0400&0.0222&0.0214&0.0127\cr
    &&MF&0.0001&0.0001&0.0013&0.0050&0.0400&0.0556&0.0500&0.0298\cr
    &&Our attack&{\bf 0.1267}&{\bf 0.1273}&{\bf 0.1283}&{\bf 0.1290}&{\bf 0.3800}&{\bf0.3444}&{\bf 0.2357}&{\bf 0.2340}\cr
     \cline{2-11}
    \hline
    \hline
    \end{tabular}
\end{table*}

For each kind of poisoning attack, we generate certain amount of fake users and extract the same number of normal users from the original dataset to form a user set. The corresponding features for each user in the user set is calculated to constitute a training dataset for fake user classifier. In our experiments, 300 normal users and 300 fake users are included in the training dataset. We follow the workflow of \colorA{SVM-TIA~\cite{zhou2016svm}} method and use the grid search with 5-fold cross validation to select the best parameters for the classifier. \colorA{SVM-TIA~\cite{zhou2016svm} is one of the state-of-the-art detection methods for shilling attacks. The detection method contains two phases. In the first phase, i.e., the support vector machine (SVM) phase, an SVM classifier is used to filter out a suspicious user set that may contain both fake users and normal users. To keep the normal users in the suspicious user set, the second target item analysis (TIA) phase tries to find out target items by counting the number of maximum rating (or minimum rating under demotion attacks) of each item in the suspicious user set. Then the items whose number of maximum rating exceeds threshold $\tau$ will be regarded as target items under the assumption that attackers will always give the maximum rating to target items. The users who set target items the maximum rating will be judged as fake users while others are viewed as normal users. Here, \colorC{$\tau$ is a hyperparameter that balances between filtering out fake users and \colorC{retaining} normal users. That is, a higher $\tau$ will cause fake users with a small attack size easier to escape detection, while a lower $\tau$ makes the detector more likely to incorrectly filter out normal users in the suspicious user set.} As the attack size of our attack can be quite small (e.g., 0.5\%), in order to \colorC{retain as many as normal users while maintaining the ability to filter out fake users}, we set $\tau$ to $0.4\%$ \colorB{(i.e., $\tau=4$ for ML-100K)} of the total number of normal users, slightly small to the smallest attack size (i.e., $0.5\%$) in our experiments.}

Note that, before training and testing the classifier, we perform data scaling on the input data, which significantly improves the model performance in this scenario. After the classifier is trained well, we can simply deploy the classifier on the recommender system to filter input training datasets and \colorD{include these users who are predicted to be fake users into the suspicious user set}. This rating scores based detection method is designed for explicit dataset. Note that, the Music dataset is purely implicit, which means that the above features will always be 0. The detection process for implicit dataset requires other complicated techniques like semantic analysis that closely relates to the platform corpus. However, the workflow for fake user detection \blue{can be} similar to that on explicit dataset, \colorC{the main difference between them is features and the way of obtaining feature values for subsequent classifier training}. Here, we mainly conduct detection experiments \mage{with random target items} on the \colorA{ML-100K dataset.}

\noindent \textbf{Effectiveness of Fake User Detectors.} In the detection process, we focus on whether the detector can effectively detect false users and whether the detector affects the original dataset. Here we use \textit{False Positive Rate (FPR)} and \textit{False Negative Rate (FNR)} to evaluate the performance of the detector, where FPR stands for the fraction of normal users who are falsely predicted as fake users while FNR means the proportion of fake users who are falsely predicted as normal users. The detection results \colorA{including both phases of SVM-TIA on the ML-100K dataset} are shown in Table \ref{tab:detection_results}. First, the TIA phase can decrease FPR after the SVM phase, while it does not influence FNR in most cases. However, we can observe that there is an abnormal increase in FNR when attack size is 0.5\% (which is slightly larger than $\tau$). This is because some fake users have escaped detection in the SVM phase and the number of the maximum ratings of the target items is lower than $\tau$. Thus, the detector cannot identify the target items and all fake users will escape detection, which further increases the FNR.
\colorA{Second}, the fake user detectors are quite efficient in detecting fake users that are generated by the baseline attacks. As we can see, FPRs and FNRs for these attacks under different attack sizes are \colorA{lower than $5\%$} in most cases, which means most fake users and normal users are correctly classified by the detectors, showing the effectiveness of this detection method. \colorA{Third}, the detector for our attack is not effective enough. FPR is around $12\%$ and FNR is around $30\%$, \colorC{which means the detector still makes a large amount of false judgements for our attack and around $30\%$ of fake users are  successfully inserted to the training dataset}. According to the above observations, it is obvious that it is much harder to detect our attack than other baseline attacks.

\noindent \textbf{Effectiveness of Poisoning Attacks under Detection.} Now we test the hit ratio of target items for different poisoning attacks after deploying fake user \blue{detectors} on the target recommender systems. The experimental results are shown in Table \ref{tab:attacks_under_detection}, where ``None'' means there is neither poisoning attack nor fake user detector deployed on the target recommender system. The hit ratio of target items with baseline attacks does not significantly change with different attack sizes. As shown in Table \ref{tab:attacks_under_detection}, \blue{overall our attack still outperforms the baseline attacks, and the baseline attacks achieve only small improvements on the initial hit ratio}. 
In particular, our attack is still effective under detection, e.g., when inserting $5\%$ fake users into the target recommender system, the hit ratio for target items rise to 0.0067, about 2.7 times of the initial hit ratio. The reason is that almost 30\% of fake users are not filtered out and they can still have a large impact on the target recommender system.
Note that, \colorC{when the attack size is small (e.g., $0.5\%$), many normal users that have rated the target items are falsely filtered out by the detector, while only few fake users are successfully injected into the dataset in our attack, which leads to relatively low hit ratios.} Even though, our attack achieves similar performance to the baseline attacks when the attack size is small.

\begin{table}[t]
\centering
  \fontsize{8}{11}\selectfont
  \caption{HR@10 for different attacks under detection.}
  \label{tab:attacks_under_detection}
    \begin{tabular}{c|c|cccc}
    \hline
    \multirow{1}{*}{Dataset}&
    \multirow{1}{*}{Attack}&
    \multicolumn{4}{c}{Attack size}\cr\cline{3-6}
    &&0.5\%&1\%&3\%&5\%\cr
    \hline
    \hline
    \multirow{5}{*}{ML-100K}&None&0.0025&0.0025&0.0025&0.0025\cr
    &Random&0.0031&0.0029&0.0023&0.0020\cr
    &Bandwagon&{\bf0.0032}&0.0029&0.0037&0.0019\cr
    &MF&0.0030&0.0029&0.0036&0.0031\cr
    &Our attack&0.0030&{\bf 0.0029}&{\bf 0.0045}&{\bf 0.0067}\cr
   
    \hline
    \hline
    \end{tabular}
\end{table}

\noindent \colorA{\textbf{Discussion.} Attackers can use various strategies to evade detection. For example, as the SVM-TIA detection method heavily relies on the frequency distribution of items, an attacker could evade detection by adjusting the process of constructing fake users, e.g., avoiding selecting the same items frequently by controlling the selection probability. Moreover, an attacker can construct  fake  users  without target items selected by default, thus decreasing the frequency of the target items. As our experimental results in Section~\ref{sec:experiments} showed,   when the target items are not selected by fake users by default, our attack remains effective and still significantly outperforms the baseline attacks.} 

{Besides the above statistical analysis of the rating patterns of normal and fake users, there are also some other detection and defense mechanisms against data poisoning attacks. For instance, Steinhardt et al.~ \cite{steinhardt2017certified} bound the training loss when the poisoned training examples are in a particular set, i.e., poisoned training examples are constrained. It is an interesting topic for future work to generalize such analysis to bound the training loss of recommender systems when an attacker can inject a bounded number of fake users. Paudice et al.~\cite{paudice2018detection} aim to statistically analyze the features of training examples and use anomaly detection to detect poisoned training examples. We explore supervised learning based defenses in our experiments, where the features are extracted from users' rating scores. As future work, we can extend the anomaly detection method to detect fake users based on statistical patterns of their rating scores.}

{There are also certifiably robust defenses~\cite{ma2019data,wang2020certifying,jia2020intrinsic,jia2020certified} against data poisoning attacks to machine learning algorithms. 
However, recommender systems are different from the machine learning algorithms considered in these work. For instance, top-$K$ items are recommended to each user in a recommender system, while a machine learning classifier predicts a single label in these work. However, it is an interesting future work to generalize these certified robustness guarantee to recommender systems.}

\section{Conclusion and Future Work}

In this work, we show that data poisoning attack to deep learning based recommender systems can be formulated as an optimization problem, which can be approximately solved via combining multiple heuristics. 
Our empirical evaluation results on three real-world datasets with different sizes show that 1) our attack can effectively promote attacker-chosen target items to be recommended to substantially more normal users, 2) our attack outperforms existing attacks, 3) our attack is still effective even if the attacker does not have access to the neural network architecture of the target recommender system and only has access to a partial user-item interaction matrix, and 4) our attack is still effective and outperforms existing attacks even if a rating score based detector is deployed. 
Interesting future work includes developing new methods to detect the fake users  and designing new recommender systems that are more robust against data poisoning attacks. 

\section*{Acknowledgement}
We thank our shepherd Jason Xue and the anonymous reviewers for their constructive comments. This work is supported in part by NSFC under Grant 61572278 and BNRist under Grant BNR2020RC01013. Qi Li is the corresponding author of this paper. 



%

\bibliographystyle{IEEEtranS}
\balance
\bibliography{reference}

\begin{thebibliography}{10}
\providecommand{\url}[1]{#1}
\csname url@samestyle\endcsname
\providecommand{\newblock}{\relax}
\providecommand{\bibinfo}[2]{#2}
\providecommand{\BIBentrySTDinterwordspacing}{\spaceskip=0pt\relax}
\providecommand{\BIBentryALTinterwordstretchfactor}{4}
\providecommand{\BIBentryALTinterwordspacing}{\spaceskip=\fontdimen2\font plus
\BIBentryALTinterwordstretchfactor\fontdimen3\font minus
  \fontdimen4\font\relax}
\providecommand{\BIBforeignlanguage}[2]{{%
\expandafter\ifx\csname l@#1\endcsname\relax
\typeout{** WARNING: IEEEtranS.bst: No hyphenation pattern has been}%
\typeout{** loaded for the language `#1'. Using the pattern for}%
\typeout{** the default language instead.}%
\else
\language=\csname l@#1\endcsname
\fi
#2}}
\providecommand{\BIBdecl}{\relax}
\BIBdecl

\bibitem{calandrino2011you}
J.~A. Calandrino, A.~Kilzer, A.~Narayanan, E.~W. Felten, and V.~Shmatikov, ``"
  you might also like:" privacy risks of collaborative filtering,'' in
  \emph{2011 IEEE Symposium on Security and Privacy}.\hskip 1em plus 0.5em
  minus 0.4em\relax IEEE, 2011, pp. 231--246.

\bibitem{Cantador:RecSys2011}
I.~Cantador, P.~Brusilovsky, and T.~Kuflik, ``2nd workshop on information
  heterogeneity and fusion in recommender systems (hetrec 2011),'' in
  \emph{Proceedings of the 5th ACM conference on Recommender systems}, ser.
  RecSys 2011.\hskip 1em plus 0.5em minus 0.4em\relax New York, NY, USA: ACM,
  2011.

\bibitem{cao2014uncovering}
Q.~Cao, X.~Yang, J.~Yu, and C.~Palow, ``Uncovering large groups of active
  malicious accounts in online social networks,'' in \emph{CCS}, 2014.

\bibitem{chen2020data}
L.~{Chen}, Y.~{Xu}, F.~{Xie}, M.~{Huang}, and Z.~{Zheng}, ``Data poisoning
  attacks on neighborhood‐based recommender systems,'' \emph{Transactions on
  Emerging Telecommunications Technologies}, 2020.

\bibitem{chen2012marginalized}
M.~Chen, Z.~Xu, K.~Weinberger, and F.~Sha, ``Marginalized denoising
  autoencoders for domain adaptation,'' \emph{arXiv preprint arXiv:1206.4683},
  2012.

\bibitem{cheng2016wide}
H.-T. Cheng, L.~Koc, J.~Harmsen, T.~Shaked, T.~Chandra, H.~Aradhye,
  G.~Anderson, G.~Corrado, W.~Chai, M.~Ispir \emph{et~al.}, ``Wide \& deep
  learning for recommender systems,'' in \emph{Proceedings of the 1st workshop
  on deep learning for recommender systems}.\hskip 1em plus 0.5em minus
  0.4em\relax ACM, 2016, pp. 7--10.

\bibitem{chirita2005preventing}
P.-A. Chirita, W.~Nejdl, and C.~Zamfir, ``Preventing shilling attacks in online
  recommender systems,'' in \emph{Proceedings of the 7th annual ACM
  international workshop on Web information and data management}.\hskip 1em
  plus 0.5em minus 0.4em\relax ACM, 2005, pp. 67--74.

\bibitem{covington2016deep}
P.~Covington, J.~Adams, and E.~Sargin, ``Deep neural networks for youtube
  recommendations,'' in \emph{Proceedings of the 10th ACM conference on
  recommender systems}.\hskip 1em plus 0.5em minus 0.4em\relax ACM, 2016, pp.
  191--198.

\bibitem{danezis2009sybilinfer}
G.~Danezis and P.~Mittal, ``Sybilinfer: Detecting sybil nodes using social
  networks.'' in \emph{NDSS}, 2009.

\bibitem{davidson2010youtube}
J.~Davidson, B.~Liebald, J.~Liu, P.~Nandy, T.~Van~Vleet, U.~Gargi, S.~Gupta,
  Y.~He, M.~Lambert, B.~Livingston \emph{et~al.}, ``The youtube video
  recommendation system,'' in \emph{Proceedings of the fourth ACM conference on
  Recommender systems}.\hskip 1em plus 0.5em minus 0.4em\relax ACM, 2010, pp.
  293--296.

\bibitem{dong2020trust}
M.~Dong, F.~Yuan, L.~Yao, X.~Wang, X.~Xu, and L.~Zhu, ``Trust in recommender
  systems: A deep learning perspective,'' \emph{arXiv preprint
  arXiv:2004.03774}, 2020.

\bibitem{fang2020influence}
M.~Fang, N.~Z. Gong, and J.~Liu, ``Influence function based data poisoning
  attacks to top-n recommender systems,'' in \emph{Proceedings of The Web
  Conference 2020}, 2020, pp. 3019--3025.

\bibitem{fang2018poisoning}
M.~Fang, G.~Yang, N.~Z. Gong, and J.~Liu, ``Poisoning attacks to graph-based
  recommender systems,'' in \emph{Proceedings of the 34th Annual Computer
  Security Applications Conference}.\hskip 1em plus 0.5em minus 0.4em\relax
  ACM, 2018, pp. 381--392.

\bibitem{fouss2007random}
F.~Fouss, A.~Pirotte, J.-M. Renders, and M.~Saerens, ``Random-walk computation
  of similarities between nodes of a graph with application to collaborative
  recommendation,'' \emph{IEEE Transactions on knowledge and data engineering},
  vol.~19, no.~3, pp. 355--369, 2007.

\bibitem{glorot2011deep}
X.~Glorot, A.~Bordes, and Y.~Bengio, ``Deep sparse rectifier neural networks,''
  in \emph{Proceedings of the fourteenth international conference on artificial
  intelligence and statistics}, 2011, pp. 315--323.

\bibitem{gong2014sybilbelief}
N.~Z. Gong, M.~Frank, and P.~Mittal, ``Sybilbelief: A semi-supervised learning
  approach for structure-based sybil detection,'' \emph{TIFS}, 2014.

\bibitem{goodfellow2014generative}
I.~Goodfellow, J.~Pouget-Abadie, M.~Mirza, B.~Xu, D.~Warde-Farley, S.~Ozair,
  A.~Courville, and Y.~Bengio, ``Generative adversarial nets,'' in
  \emph{Advances in neural information processing systems}, 2014, pp.
  2672--2680.

\bibitem{gunes2014shilling}
I.~Gunes, C.~Kaleli, A.~Bilge, and H.~Polat, ``Shilling attacks against
  recommender systems: a comprehensive survey,'' \emph{Artificial Intelligence
  Review}, vol.~42, no.~4, pp. 767--799, 2014.

\bibitem{harper2015movielens}
F.~M. Harper and J.~A. Konstan, ``The movielens datasets: History and
  context,'' \emph{Acm transactions on interactive intelligent systems (tiis)},
  vol.~5, no.~4, pp. 1--19, 2015.

\bibitem{he2017neural}
X.~He, L.~Liao, H.~Zhang, L.~Nie, X.~Hu, and T.-S. Chua, ``Neural collaborative
  filtering,'' in \emph{Proceedings of the 26th international conference on
  world wide web}.\hskip 1em plus 0.5em minus 0.4em\relax International World
  Wide Web Conferences Steering Committee, 2017, pp. 173--182.

\bibitem{hornik1989multilayer}
K.~Hornik, M.~Stinchcombe, and H.~White, ``Multilayer feedforward networks are
  universal approximators,'' \emph{Neural networks}, vol.~2, no.~5, pp.
  359--366, 1989.

\bibitem{jia2020certified}
J.~Jia, X.~Cao, and N.~Z. Gong, ``Certified robustness of nearest neighbors
  against data poisoning attacks,'' \emph{Arxiv}, 2020.

\bibitem{jia2020intrinsic}
------, ``Intrinsic certified robustness of bagging against data poisoning
  attacks,'' in \emph{AAAI}, 2021.

\bibitem{jia2017random}
J.~Jia, B.~Wang, and N.~Z. Gong, ``Random walk based fake account detection in
  online social networks,'' in \emph{DSN}, 2017.

\bibitem{kapoor2017review}
S.~Kapoor, V.~Kapoor, and R.~Kumar, ``A review of attacks and its detection
  attributes on collaborative recommender systems.'' \emph{International
  Journal of Advanced Research in Computer Science}, vol.~8, no.~7, 2017.

\bibitem{koren2009matrix}
Y.~Koren, R.~Bell, and C.~Volinsky, ``Matrix factorization techniques for
  recommender systems,'' \emph{Computer}, no.~8, pp. 30--37, 2009.

\bibitem{lam2004shilling}
S.~K. Lam and J.~Riedl, ``Shilling recommender systems for fun and profit,'' in
  \emph{Proceedings of the 13th international conference on World Wide
  Web}.\hskip 1em plus 0.5em minus 0.4em\relax ACM, 2004, pp. 393--402.

\bibitem{li2016data}
B.~Li, Y.~Wang, A.~Singh, and Y.~Vorobeychik, ``Data poisoning attacks on
  factorization-based collaborative filtering,'' in \emph{Advances in neural
  information processing systems}, 2016, pp. 1885--1893.

\bibitem{ma2019data}
Y.~{Ma}, X.~{Zhu}, and J.~{Hsu}, ``Data poisoning against
  differentially-private learners: Attacks and defenses,'' in \emph{Proceedings
  of the Twenty-Eighth International Joint Conference on Artificial
  Intelligence}, 2019, pp. 4732--4738.

\bibitem{mnih2015human}
V.~Mnih, K.~Kavukcuoglu, D.~Silver, A.~A. Rusu, J.~Veness, M.~G. Bellemare,
  A.~Graves, M.~Riedmiller, A.~K. Fidjeland, G.~Ostrovski \emph{et~al.},
  ``Human-level control through deep reinforcement learning,'' \emph{Nature},
  vol. 518, no. 7540, p. 529, 2015.

\bibitem{mobasher2007attacks}
B.~Mobasher, R.~Burke, R.~Bhaumik, and J.~J. Sandvig, ``Attacks and remedies in
  collaborative recommendation,'' \emph{IEEE Intelligent Systems}, vol.~22,
  no.~3, pp. 56--63, 2007.

\bibitem{mobasher2007toward}
B.~Mobasher, R.~Burke, R.~Bhaumik, and C.~Williams, ``Toward trustworthy
  recommender systems: An analysis of attack models and algorithm robustness,''
  \emph{ACM Transactions on Internet Technology (TOIT)}, vol.~7, no.~4, p.~23,
  2007.

\bibitem{munoz2017towards}
L.~Mu{\~n}oz-Gonz{\'a}lez, B.~Biggio, A.~Demontis, A.~Paudice, V.~Wongrassamee,
  E.~C. Lupu, and F.~Roli, ``Towards poisoning of deep learning algorithms with
  back-gradient optimization,'' in \emph{Proceedings of the 10th ACM Workshop
  on Artificial Intelligence and Security}.\hskip 1em plus 0.5em minus
  0.4em\relax ACM, 2017, pp. 27--38.

\bibitem{okura2017embedding}
S.~Okura, Y.~Tagami, S.~Ono, and A.~Tajima, ``Embedding-based news
  recommendation for millions of users,'' in \emph{Proceedings of the 23rd ACM
  SIGKDD International Conference on Knowledge Discovery and Data
  Mining}.\hskip 1em plus 0.5em minus 0.4em\relax ACM, 2017, pp. 1933--1942.

\bibitem{o2005recommender}
M.~P. O'Mahony, N.~J. Hurley, and G.~C. Silvestre, ``Recommender systems:
  Attack types and strategies,'' in \emph{AAAI}, 2005, pp. 334--339.

\bibitem{paudice2018detection}
A.~{Paudice}, L.~{Muñoz-González}, A.~{György}, and E.~C. {Lupu},
  ``Detection of adversarial training examples in poisoning attacks through
  anomaly detection.'' \emph{arXiv preprint arXiv:1802.03041}, 2018.

\bibitem{pi2018survey}
H.~Pi, Z.~Ji, and C.~Yang, ``A survey of recommender system from data sources
  perspective,'' in \emph{2018 8th International Conference on Management,
  Education and Information (MEICI 2018)}.\hskip 1em plus 0.5em minus
  0.4em\relax Atlantis Press, 2018.

\bibitem{sarwar2001item}
B.~Sarwar, G.~Karypis, J.~Konstan, and J.~Riedl, ``Item-based collaborative
  filtering recommendation algorithms,'' in \emph{Proceedings of the 10th
  international conference on World Wide Web}, 2001, pp. 285--295.

\bibitem{steinhardt2017certified}
J.~{Steinhardt}, P.~W.~W. {Koh}, and P.~S. {Liang}, ``Certified defenses for
  data poisoning attacks,'' in \emph{Proceedings of the 31st International
  Conference on Neural Information Processing Systems}, 2017, pp. 3517--3529.

\bibitem{wang2020certifying}
B.~Wang, X.~Cao, J.~Jia, and N.~Z. Gong, ``On certifying robustness against
  backdoor attacks via randomized smoothing,'' in \emph{CVPR 2020 Workshop on
  Adversarial Machine Learning in Computer Vision}, 2020.

\bibitem{wang2017gang}
B.~Wang, N.~Z. Gong, and H.~Fu, ``Gang: Detecting fraudulent users in online
  social networks via guilt-by-association on directed graphs,'' in
  \emph{ICDM}, 2017.

\bibitem{wang19ndss}
B.~Wang, J.~Jia, and N.~Z. Gong, ``Graph-based security and privacy analytics
  via collective classification with joint weight learning and propagation,''
  in \emph{NDSS}, 2019.

\bibitem{wang2017sybilscar}
B.~Wang, L.~Zhang, and N.~Z. Gong, ``Sybilscar: Sybil detection in online
  social networks via local rule based propagation,'' in \emph{INFOCOM}, 2017.

\bibitem{wang2013you}
G.~Wang, T.~Konolige, C.~Wilson, X.~Wang, H.~Zheng, and B.~Y. Zhao, ``You are
  how you click: Clickstream analysis for sybil detection,'' in \emph{USENIX
  Security}, 2013.

\bibitem{xing2013take}
X.~Xing, W.~Meng, D.~Doozan, A.~C. Snoeren, N.~Feamster, and W.~Lee, ``Take
  this personally: Pollution attacks on personalized services,'' in
  \emph{Presented as part of the 22nd $\{$USENIX$\}$ Security Symposium
  ($\{$USENIX$\}$ Security 13)}, 2013, pp. 671--686.

\bibitem{yang2017fake}
G.~Yang, N.~Z. Gong, and Y.~Cai, ``Fake co-visitation injection attacks to
  recommender systems.'' in \emph{NDSS}, 2017.

\bibitem{yu2006sybilguard}
H.~Yu, H.~Yu, M.~Kaminsky, P.~B. Gibbons, and A.~Flaxman, ``Sybilguard:
  defending against sybil attacks via social networks,'' in \emph{SIGCOMM},
  2006.

\bibitem{yuan2019detecting}
D.~Yuan, Y.~Miao, N.~Z. Gong, Z.~Yang, Q.~Li, D.~Song, Q.~Wang, and X.~Liang,
  ``Detecting fake accounts in online social networks at the time of
  registrations,'' in \emph{CCS}, 2019.

\bibitem{zeller2008cross}
W.~Zeller and E.~W. Felten, ``Cross-site request forgeries: Exploitation and
  prevention,'' \emph{The New York Times}, pp. 1--13, 2008.

\bibitem{zhang2019deep}
S.~Zhang, L.~Yao, A.~Sun, and Y.~Tay, ``Deep learning based recommender system:
  A survey and new perspectives,'' \emph{ACM Computing Surveys (CSUR)},
  vol.~52, no.~1, p.~5, 2019.

\bibitem{zhou2016svm}
W.~Zhou, J.~Wen, Q.~Xiong, M.~Gao, and J.~Zeng, ``Svm-tia a shilling attack
  detection method based on svm and target item analysis in recommender
  systems,'' \emph{Neurocomputing}, vol. 210, pp. 197--205, 2016.

\end{thebibliography}

\appendix

\begin{table*}[t]
\centering
  \fontsize{8}{11}\selectfont
  \caption{Standard deviations for different attacks with different attack sizes.}
  \label{tab:std_attack_size}
    \begin{tabular}{c|c|cccc|cccc}
    \hline
    \multirow{3}{*}{Dataset}&
    \multirow{3}{*}{Attack}&
    \multicolumn{8}{c}{Attack size}\cr\cline{3-10}
    &&\multicolumn{4}{c|}{Random target items}&\multicolumn{4}{c}{Unpopular target items}\cr\cline{3-10}
    &&0.5\%&1\%&3\%&5\%&0.5\%&1\%&3\%&5\%\cr
    \hline
    \hline
    \multirow{5}{*}{ML-100K}&None&0.0033&0.0033&0.0033&0.0033&0&0&0&0\cr
    &Random&0.0036&0.0041&0.0054&0.0074&0.0002&0.0003&0.0009&0.0016\cr
    &Bandwagon&{\bf 0.0038}&0.0039&0.0056&0.0075&0.0002&0.0003&0.0009&0.0015\cr
    &MF&{\bf 0.0038}&{\bf 0.0045}&0.0070&0.0089&0.0002&0.0003&0.0020&0.0032\cr
    &Our attack&{\bf 0.0038}&0.0043&{\bf 0.0090}&{\bf 0.0122}&{\bf 0.0008}&{\bf 0.0021}&{\bf 0.0060}&{\bf 0.0101}\cr
    \hline
    \multirow{5}{*}{Music}&None&0.0038&0.0038&0.0038&0.0038&0.0005&0.0005&0.0005&0.0005\cr
    &Random&0.0044&0.0049&0.0063&0.0098&0.0008&0.0015&0.0041&0.0064\cr
    &Bandwagon&0.0044&0.0049&0.0068&0.0076&0.0006& 0.0014&0.0029&0.0061\cr
    &MF&0.0043&0.0052&0.0073&0.0084&0.0005&0.0018&0.0045&0.0062\cr
    &Our attack&{\bf 0.0055}&{\bf 0.0066}&{\bf 0.0079}&{\bf 0.0109}&{\bf 0.0015}&{\bf 0.0025}&{\bf 0.0063}&{\bf 0.0100}\cr
    \hline
    \hline
    \end{tabular}
\end{table*}

\subsection{Standard Deviations of Experimental Results}\label{sec:standard_diviations}
\colorC{In this section, we provide the standard deviations of experimental results (see  Table \ref{tab:attack_size_comparison} in Section \ref{sec:effect_poison}), which is corresponding to that in Table \ref{tab:std_attack_size}. We can observe some interesting findings in Table \ref{tab:attack_size_comparison} and Table \ref{tab:std_attack_size}. First, the increase of standard deviations is slower than that of the average hit ratios. For example, on the ML-100K dataset, the average hit ratio for random target items is 0.0025 for the ``None" setting, while the standard deviation of the hit ratios for these target items is 0.0033, even larger than the former. As for all attack methods, after injecting $5\%$ fake users into the target recommender system, all average hit ratios for random target items are larger than the standard deviations for hit ratios of these target items, which indicates that all attack methods can promote target items. Second, our attack have the highest standard deviations in most cases. The reason is that our attack promotes target items most significantly among all attacks and the hit ratios for some target items tend to increase faster than others.}

\subsection{Hit Ratio per Target Item}\label{sec:HR_per_target}

\colorC{In this section, we show the change of the hit ratio for each target item in different attacks with different attack sizes. We count the number of target items whose hit ratio has been promoted compared to the original value. The results are shown in Table \ref{tab:HR_per_target}. We find that not all target items can get promoted when injecting limited number of fake users, which \colorD{often} happens when the attack size is very small. As the attack size increases, more target items get promoted and finally all target items obtain an increased hit ratio in all attack methods when $5\%$ fake users are injected. More importantly, \colorD{we observe that our attack increases the hit ratios of most selected target items, especially when the attack size is small, e.g., $0.5\%$.} All these results demonstrate our attack is effective to promote target items in poison deep learning based recommender systems.}

\begin{table*}[t]
\centering
  \fontsize{8}{11}\selectfont
  \caption{The number of promoted target items for different attacks with different attack sizes.}
  \label{tab:HR_per_target}
    \begin{tabular}{c|c|cccc|cccc}
    \hline
    \multirow{3}{*}{Dataset}&
    \multirow{3}{*}{Attack}&
    \multicolumn{8}{c}{Attack size}\cr\cline{3-10}
    &&\multicolumn{4}{c|}{Random target items}&\multicolumn{4}{c}{Unpopular target items}\cr\cline{3-10}
    &&0.5\%&1\%&3\%&5\%&0.5\%&1\%&3\%&5\%\cr
    \hline
    \hline
    \multirow{5}{*}{ML-100K}&None&0&0&0&0&0&0&0&0\cr
    &Random&4&8&9&{\bf 10}&6&9&{\bf 10}&{\bf 10}\cr
    &Bandwagon&6&8&{\bf 10}&{\bf 10}&6&{\bf 10}&{\bf 10}&{\bf 10}\cr
    &MF&7&8&{\bf 10}&{\bf 10}&5&8&{\bf 10}&{\bf 10}\cr
    &Our attack&{\bf 9}&{\bf 9}&{\bf 10}&{\bf 10}&{\bf 9}&{\bf 10}&{\bf 10}&{\bf 10}\cr
    \hline
    \multirow{5}{*}{Music}&None&0&0&0&0&0&0&0&0\cr
    &Random&9&{\bf 10}&{\bf 10}&{\bf 10}&7&9&{\bf 10}&{\bf 10}\cr
    &Bandwagon&8&{\bf 10}&{\bf 10}&{\bf 10}&8& 9&{\bf 10}&{\bf 10}\cr
    &MF&{\bf 10}&{\bf 10}&{\bf 10}&{\bf 10}&8&9&{\bf 10}&{\bf 10}\cr
    &Our attack&9&{\bf 10}&{\bf 10}&{\bf 10}&{\bf 9}&{\bf 10}&{\bf 10}&{\bf 10}\cr
    \hline
    \hline
    \end{tabular}

\end{table*}

\end{document}